\documentclass[superscriptaddress,nobibnotes,amsmath,amssymb,notitlepage,twocolumn,prx,longbibliography]{revtex4-2}

\usepackage{bm,mathptmx,braket}

\usepackage{graphicx,color,hyperref}
\usepackage[caption=false]{subfig}
\hypersetup{colorlinks=true, linkcolor=blue, citecolor=blue, urlcolor=blue} 

\graphicspath{{./img/}}

\newcommand{\beq}[1]{\begin{equation}\label{#1}}
\newcommand{\eep}{\;.\end{equation}}
\newcommand{\eec}{\;,\end{equation}}
\newcommand{\eeq}{\end{equation}}

\newcommand{\om}{\omega}

\DeclareMathAlphabet{\mathcal}{OMS}{cmsy}{m}{n} 

\begin{document}

\title{Frustrated Topological Orbital Magnetism}


\author{Sergio Garrido Bourkhaeva}

\email{sg2114@cam.ac.uk}
\affiliation{\TCM}

\author{Claudio Castelnovo}
\email{cc726@cam.ac.uk}
\affiliation{\TCM}

\author{Wojciech J. Jankowski}
\email{wjj25@cam.ac.uk}
\affiliation{\TCM}
\affiliation{\MPIPKS}

\newcommand{\TCM}{{Theory of Condensed Matter Group, Cavendish Laboratory, University of Cambridge, J.\,J.\,Thomson Avenue, Cambridge CB3 0HE, UK}}

\newcommand{\MPIPKS}{{Max-Planck-Institut f\"ur Physik komplexer Systeme, N\"othnitzer Stra\ss e 38, 01187 Dresden, Germany}}

\date{\today}

\begin{abstract}
    By studying the interplay of geometric frustration with Aharonov--Bohm fluxes in three dimensions, we unveil a rich playground exhibiting topological semimetallic ground states with orbital magnetic dipole, quadrupole and ferrotoroidic orders, akin to spin patterns in classical spin ices --- a class of phenomena which we dub frustrated topological orbital magnetism. We demonstrate that the ground-state toroidizations underpinned by non-Abelian topological nodal lines induce measurable current-induced Hall phenomena. We further find that distinct 2-in-2-out and 3-in-1-out frustrated orbital current patterns can be associated with topological Weyl node charges, and corresponding anisotropic linear and nonlinear Hall effects.~We propose signatures for \mbox{experimental} validation of distinct frustrated topological orbital magnetic orders in candidate materials.
\end{abstract}

\maketitle
%
%

\section{Introduction} 
Frustration in condensed matter systems typically arises from a competition between interactions and lattice geometry that prevents the straightforward minimization of each local energy term, thus obstructing conventional ordering~\cite{Moessner2006}. Frustrated magnetism constitutes a hallmark example of geometric frustration~\cite{Ramirez1994, Lacroix2011}. For instance, antiferromagnetic spin interactions on the kagome and pyrochlore lattices give rise to new phases of matter, known as spin liquids~\cite{Balents2010, Savary2017, Knolle2019}. 
Such phases are characterized by emergent gauge symmetries~\cite{Wen2004, Levin2005}, topological order~\cite{Wen2017}, and fractionalized excitations~\cite{Kivelson1987, Broholm2020}.

A case in point that is relevant to our discussion below are classical spin ices~\cite{Udagawa2021}. These are a class of models and materials where geometrically frustrated interactions on the pyrochlore lattice give rise to local 2-in-2-out correlations (known as ice rules) and extensive degeneracy~\cite{Harris1997, Ramirez1999, Bramwell2001} --- analogous to the patterns of hydrogen bonding between water molecules in ice~\cite{Pauling1935}. 
Spin-flip excitations in spin ices fractionalize at low energy into deconfined emergent magnetic monopoles connected by Dirac strings~\cite{Castelnovo2008, Jaubert2009, Castelnovo2012}, as experimentally demonstrated, e.g., in pyrochlore materials Ho$_2$Ti$_2$O$_7$ or Dy$_2$Ti$_2$O$_7$~\cite{Fennell2009, Morris2009, Bramwell2009, Kadowaki2009}.

\begin{figure}[t!]
\centering
  \includegraphics[width=\columnwidth]{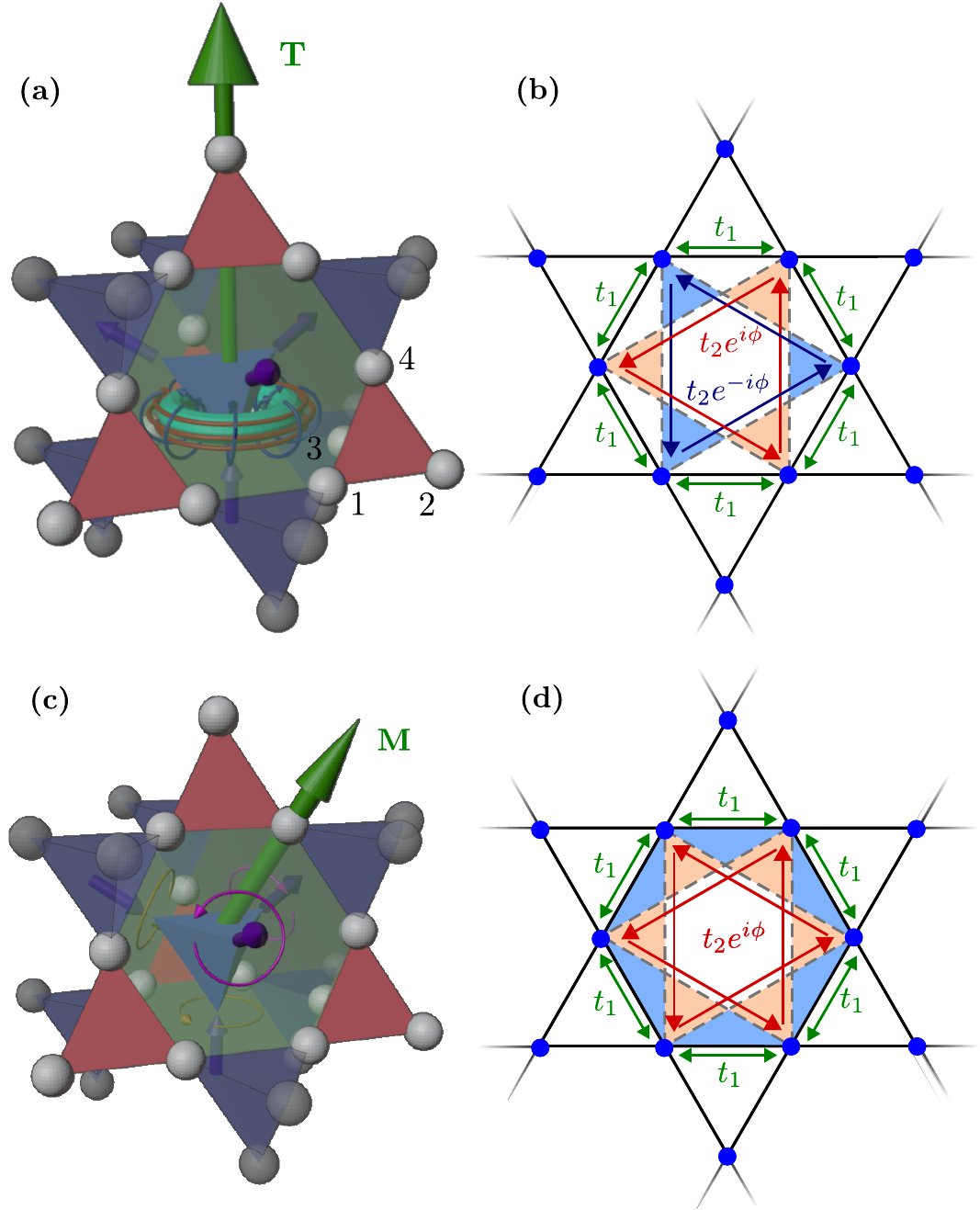}
  \caption{Frustrated topological orbital magnetic orders. {\bf (a)},~{\bf(c)} Pyrochlore lattice realizations of geometrically frustrated orbital magnetism and corresponding staggered magnetic fluxes. {\bf (b)},~{\bf(d)} Kagome planes illustrating the 
  Aharonov--Bohm phases $(\phi)$ and \mbox{tunnelling} amplitudes between nearest neighbor $(t_1)$, and next-nearest neighbor $(t_2)$ orbitals. {\bf (a)}~Frustrated toroidal orbital \mbox{magnet} with \mbox{toroidization}~$\textbf{T}$ in a 1-in-3-out (\text{dark~purple arrows}) tetrahedral configuration of antichiral orbital magnetization currents. {\bf (b)}~\mbox{Antichiral} pattern of fluxes, resulting in frustrated \mbox{topological} orbital ferrotoroidic order.{\bf (c)}~Frustrated orbital ferromagnetic \mbox{order} with Weyl semimetal topology and orbital magnetization $\textbf{M}$ in~\mbox{a 2-in-2-out} \mbox{configuration} of orbital currents. {\bf (d)}~Chiral pattern of magnetic fluxes, resulting in the~geometrically frustrated topological orbital ferromagnetic~\mbox{order}.}
\label{fig:1}
\end{figure}

While frustrated magnetism has been extensively studied in spin models, to the best of our knowledge, \mbox{analogous} \mbox{geometric} frustration of orbital magnetism, namely frustrated loop currents, has not received attention to date. In~this~work, we~identify and characterize such exotic phases of \mbox{matter}, and~we~demonstrate the key role of band topology in \mbox{realizing} geometrically frustrated orbital magnetism.~We~thus \mbox{refer} to these uncovered phases as frustrated topological \mbox{orbital magnets} (FTOMs). 

We demonstrate the interplay of geometric frustration and topological orbital magnetism in effective spinless models on the pyrochlore lattice, which we describe in detail in App.~\ref{app:models}. The lattice is tetrapartite, and we shall make use in \mbox{particular} of the fact that any three of the four sublattices form a layered kagome lattice supporting orbital loop currents, see~Fig.~\ref{fig:1}.  
We show how distinct geometrically \mbox{frustrated} \mbox{patterns} of magnetic Aharonov--Bohm fluxes, threaded through kagome planes, result in different frustrated topological orbital magnetic phases. We classify these phases by (i)~mapping to the ice rule configurations of classical spin ices, and~by (ii)~the presence of ground-state orbital magnetization and toroidization vectors. We further connect the latter to observable Hall \mbox{phenomena}, which may serve as smoking-gun signatures of~the proposed frustrated topological orbital magnetic orders. Finally, we discuss possible experimental observation protocols, 
candidate materials, and directions for further study.

The significance of the central results of our work can be summarized as follows. First, by predicting hitherto uncharted phases of matter with geometrically frustrated Aharonov--Bohm flux patterns, our results open a window to studying novel magnetic orders. Second, from a theoretical viewpoint, these unprecedented phases offer important platforms for probing lattice gauge theories and \mbox{quantum} dynamics \mbox{constrained} by geometric frustration. \mbox{Finally}, an \mbox{experimental} discovery of the proposed phases would offer great promise for unprecedented nanotechnological \mbox{applications} that utilize the exotic interplay of~dissipationless topological \mbox{currents} and frustrated magnetic properties in electronic devices \cite{Gilbert2021, He2022}.~Overall, we envision that \mbox{frustrated} topological \mbox{orbital} \mbox{magnetic} orders shall provide a~rich \mbox{platform} for new \mbox{phenomena} at the intersection of \mbox{topological} and frustrated quantum \mbox{matter}. 
%
%

\section{Classification} 
We begin by subdividing the proposed FTOM phases into two classes. The first class realizes ground state orbital toroidization~\cite{Shitade2018, Gao2018}, 
\begin{equation}\label{eq::T}
   \textbf{T} = \frac{1}{V} \int \text{d} \textbf{r}~[\textbf{r} \times \textbf{m}(\textbf{r})]
   \, , 
\end{equation}
where $V$ is the volume of the crystal, and $\textbf{m}(\textbf{r}) = \langle \textbf{r} \times \mathbf{j}(\textbf{r}) \rangle$ is the expectation value of a local orbital magnetization due to the circulating local current densities $\mathbf{j}(\textbf{r})$ at positions $\textbf{r}$. 
Frustrated toroidal orbital magnets, see Fig.~\ref{fig:1}(a), arise from alternating, antichiral, Aharonov--Bohm phase patterns, illustrated e.g., in Fig.~\ref{fig:1}(b). By construction, such patterns satisfy $\mathcal{PT}$ symmetry, i.e., spatiotemporal inversion symmetry, with $\mathcal{P}$ the spatial inversion symmetry, and $\mathcal{T}$ the time-reversal symmetry. Notably, the order parameter $\textbf{T}$ also satisfies $\mathcal{PT}$ symmetry by construction --- under inversion $\mathcal{P}: \textbf{r} \rightarrow - \textbf{r}$, while time-reversal inverts the orbital current and therefore magnetization, $\mathcal{T}: \textbf{m}(\textbf{r}) \rightarrow - \textbf{m}(\textbf{r})$. As we will show below, a~fingerprint of nonvanishing orbital toroidization in FTOMs is a current-induced zero-field Hall effect~\cite{Khomskii2014, Mineev2025}. Centrally to this work, we further demonstrate that the emergence of orbital toroidization and associated current-induced Hall effect can be traced to non-Abelian topological nodal lines protected by $\mathcal{PT}$ symmetry~\cite{Wu2019,Tiwari2020}. 
We illustrate these topological \mbox{relations} in both 2-in-2-out and 3-in-1-out orbital current magnetization vector configurations in toroidal magnets. 

The second class of frustrated topological orbital magnetic ground states realizes net orbital magnetization~\cite{Ceresoli2006, Xiao2010}, which vanishes in toroidal magnets, 
\begin{equation}
    \textbf{M} = \frac{1}{V} \int \text{d} \textbf{r}~\textbf{m}(\textbf{r})
    \, . 
\end{equation}
Within the pyrochlore lattice, $\textbf{M}$ is supported by loop \mbox{currents} on distinct kagome planes, see Fig.~\ref{fig:1}(c). As we shall \mbox{discuss}, these are induced by 2-in-2-out or 3-in-1-out configurations of~orbital loop currents imposed by in-phase Aharonov--Bohm fluxes on each individual 
kagome plane, 
see Fig.~\ref{fig:1}(d).
We~demonstrate that nonvanishing orbital magnetizations derive from the band topology of Weyl nodes, acting as Berry curvature monopoles, and from Chern bands,
in contrast to the toroidal magnets. 
The frustrated Aharonov--Bohm fluxes in this case respect $\mathcal{P}$ symmetry while breaking $\mathcal{T}$ symmetry, i.e., the systems are centrosymmetric magnets by construction. Correspondingly, the 2-in-2-out or 3-in-1-out configurations realize distinct Berry curvature profiles and \mbox{different} anisotropic anomalous Hall effects, which allow one to map out the underlying frustrated orbital magnetic order \mbox{experimentally}. 

Throughout this work, we focus on spinless \mbox{lattice} \mbox{models}, demonstrating that geometrically frustrated \mbox{topological} \mbox{orbital} magnetic orders can be realized in the limit of \mbox{vanishing} coupling between the spin and orbital degrees of freedom. 
Correspondingly, the Weyl semimetallic FTOM phases uncovered in our work have a notably different \mbox{origin} with respect to the ones already studied in spinful models on~the pyrochlore lattice~\cite{Wan2011, Berke2018}. 
%

\section{Frustrated toroidal orbital magnetism}
\begin{figure*}[t!]
\centering
  \includegraphics[width=\linewidth]{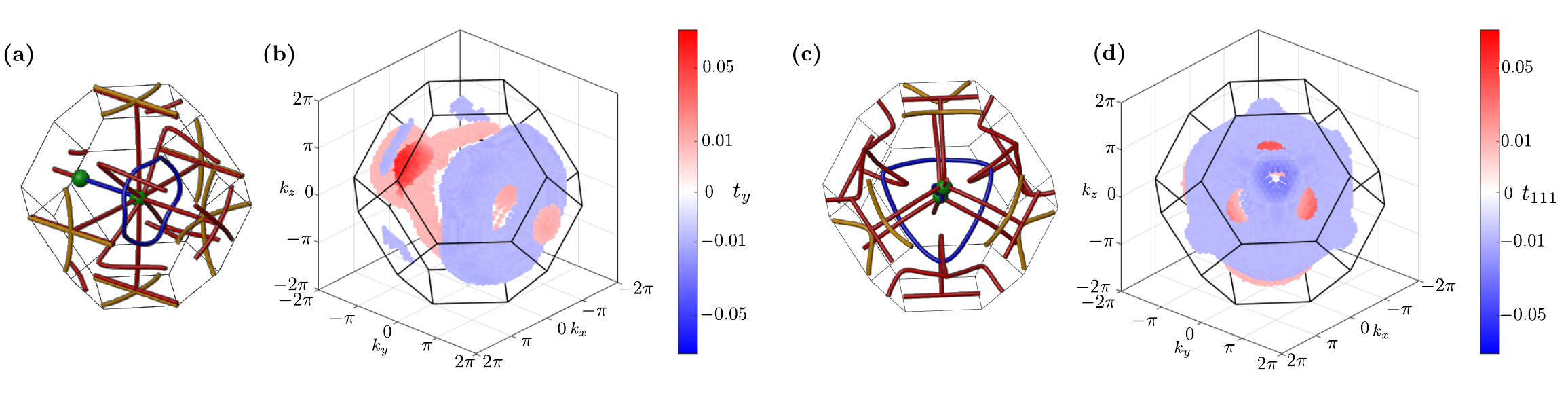}
  \caption{Frustrated orbital toroidization from $\mathcal{PT}$-symmetric non-Abelian topological nodal lines. {\bf (a)},~{\bf (c)}  Topological nodal lines \mbox{underpinning} the toroidization through non-Abelian Berry connection vortices in 2-in-2-out and 3-in-1-out flux configurations, respectively. {\bf (b)},~{\bf (d)} Momentum-space-resolved toroidization densities contributed by single-particle states supporting the nodal structures over the~BZ, in~\mbox{2-in-2-out} and 3-in-1-out antichiral Aharonov--Bohm flux configurations. The nodal lines between bands $n=1,2$, $n=2,3$, and $n=3,4$ are shown in red, blue, and yellow, respectively. The green spheres represent the triple nodal points.}
\label{fig:2}
\end{figure*}
We first consider the effective models of frustrated toroidal orbital magnets. The corresponding staggered fluxes through the kagome planes of the pyrochlore lattice are analogous to those of a modified Haldane model with antichiral edge states~\cite{Franz2018Antichiral}, see Fig.~\ref{fig:1}(b). These systems break both $\mathcal{P}$- and $\mathcal{T}$-symmetry independently while preserving $\mathcal{PT}$-symmetry, 
thus allowing 
toroidal order to develop. 

We introduce four orbital sites within each up-tetrahedron of the pyrochlore lattice, shown in red in Fig.~\ref{fig:1}, with position vectors $ \mathbf{r}_1 = \frac{1}{4} (0,0,0)$, $\mathbf{r}_2 = \frac{1}{4} (1,1,0)$, $\mathbf{r}_3 = \frac{1}{4} (0,1,1)$ and $\mathbf{r}_4 = \frac{1}{4} (1,0,1)$, as we detail in App.~\ref{app:models}, Fig.~\ref{fig:6}. We define displacement vectors ${\mathbf{r}_{ij} = \mathbf{r}_{j} - \mathbf{r}_{i}}$, with ${i,j=1,2,3,4}$. The Bloch Hamiltonian for a single particle with crystal momentum $\textbf{k}$ in the Brillouin zone (BZ) and nearest neighbor (NN) hopping amplitude $t_1$ can then be  written in terms of hopping matrix elements ${\mathcal{H}_{ij}^{\text{\text{NN}}}(\textbf{k})= - 2 t_{1} \cos (\mathbf{k} \cdot \mathbf{r}_{ij} )}$, with $i \neq j$. 

We add next nearest neighbor (NNN) hopping with amplitude $t_2$ within the kagome planes. Each hexagon in the kagome planes is threaded by a pattern of staggered fluxes that couple to the particle hopping as Aharonov--Bohm phases~$\phi$, $t_2 e^{\text{i}\phi}$, as~per standard Peierls substitution. We label the fluxes through the hexagons as $\phi_{123}$, $\phi_{124}$, $\phi_{134}$, and $\phi_{234}$, see App.~\ref{app:models}, Fig.~\ref{fig:6}, where the indices denote the sites belonging to the \mbox{corresponding} kagome planes. 
 
 As a reference configuration, we define an `all-in' (\mbox{4-in-0-out}) pattern as one where all fluxes are equal, ${(\phi_{123}, \phi_{124}, \phi_{134}, \phi_{234})=(\phi, \phi, \phi, \phi)}$. Accordingly, a 2-in-2-out pattern has ${(\phi_{123}, \phi_{124}, \phi_{134}, \phi_{234})=(\phi, \phi, -\phi, -\phi)}$. 
\mbox{Under} the antichiral pattern of fluxes, i.e., an alternating~$\pm\phi$ Aharonov--Bohm phase pattern between the second-neighbor hoppings across honeycomb hexagons, see Fig.~\ref{fig:1}(b), the single particle Bloch Hamiltonian for instance between sites ${i=1}$ and ${j=2}$ can be written as $\mathcal{H}_{12}^{\text{\text{NNN}}}(\textbf{k}) = - 2 t_{2} [ {\cos (\mathbf{k} \cdot (\mathbf{r}_{13} + \mathbf{r}_{23}) + \phi_{123})} + {\cos (\mathbf{k} \cdot (\mathbf{r}_{14} + \mathbf{r}_{24}) + \phi_{124})]}$.

We find that the class of Hamiltonians obtained by combining $\mathcal{H}_{ij}^{\text{\text{NN}}}$ and $\mathcal{H}_{ij}^{\text{\text{NNN}}}$, see App.~\ref{app:models}, hosts topological nodal lines in their \mbox{energy} spectra. In Fig.~\ref{fig:2}, we show that these are closely related to the frustrated \mbox{orbital} magnetism of the corresponding ground states. First, we note that the $\mathcal{PT}$-symmetric spinless systems admit a gauge in which all Bloch eigenstates $\ket{u_{n\textbf{k}}}$ can be chosen real~\cite{BzdusekTP}. In this gauge, the nodal lines can be assigned a~topological charge through computation of the integral around the nodes, 
\begin{equation}
    \phi_{nm} = \oint \text{d}\textbf{k} \cdot\textbf{A}_{nm} \in \pi \mathbb{Z}
    \, , 
\end{equation}
with the non-Abelian Berry connection, $\textbf{A}_{nm} = \braket{u_{n\textbf{k}}|\boldsymbol{\nabla}_\textbf{k} u_{m\textbf{k}}}$, between~the corresponding bands $n,m$. 

In Fig.~\ref{fig:2}(a) and Fig.~\ref{fig:2}(c), we show the network of nodal lines with nonzero topological charge ($\phi_{nm} = \pi$) for both a~\mbox{2-in-2-out} and a {1-in-3-out} pattern with fluxes $\phi=\frac{\pi}{2}$. 
We find that the nodal lines passing through the $\Gamma$ point (\mbox{BZ center}) along rotational $\mathcal{C}_{3}$/$\mathcal{C}_{4}$ axes, terminate in nodal triple points, where three band energies cross. The topological triple points, within the classification of Ref.~\onlinecite{BzdusekTP}, are of type~A for \mbox{2-in-2-out} configurations with a $\mathcal{C}_{4}$-symmetry, and of type~B, for \mbox{1-in-3-out} configurations with $\mathcal{C}_{3}$-symmetry.

We then compute the orbital toroidization using a multiband generalization of the expression for the orbital magnetic quadrupole moment, $\mathcal{Q}_{ij}$~\cite{GaoQ}, which we derive within a band projector formalism in App.~\ref{app:projector}. $T_{k} = \frac{1}{2} \varepsilon_{ijk} \mathcal{Q}_{ij}$, as the toroidization equals the antisymmetric part of the magnetic quadrupole moment $\mathcal{Q}_{ij}$ 
--- extracted here using the totally \mbox{antisymmetric} Levi-Civita tensor $\varepsilon_{ijk}$,
as one can see from the definition,~Eq.~\eqref{eq::T}. 

In Fig.~\ref{fig:2}(b) and Fig.~\ref{fig:2}(d), we plot the toroidization density projected on individual $\textbf{k}$-points, i.e., the single-particle contributions at different momenta, along the direction of the net toroidization, for both a 2-in-2-out and a 1-in-3-out configuration with $\phi=\frac{\pi}{2}$, and chemical potential $\mu$ that maximizes the net toroidization $\textbf{T}$.
We find the toroidization density to concentrate in concentric toroidal structures around nodal lines with triple nodal points. The positive and negative contributions are uneven, consistent with the broken $\mathcal{P}: \textbf{k} \rightarrow -\textbf{k}$ \mbox{symmetry}, as we detail in App.~\ref{app:projector}. 

Our results demonstrate how the orbital toroidization~$\textbf{T}$ of FTOMs is underpinned by the rich geometry of the non-Abelian Berry connection $\textbf{A}_{nm}$ around the topological nodal lines, which enters the computation of $\mathcal{Q}_{ij}$. This is further \mbox{evidenced} by the change in sign of the toroidization density~$\textbf{t}(\textbf{k})$ upon traversing the blue nodal loop which encircles the nodal line with triple nodal points in both Fig.~\ref{fig:2}(b) and Fig.~\ref{fig:2}(d). Moreover, the interplay of $\textbf{t}(\textbf{k})$ with non-Abelian band geometry is manifested by the peak toroidization density around the nodal lines emanating from the triple nodal points, which is larger than at the other $\textbf{k}$-points related by a~$\frac{\pi}{3}$ rotation around the $[111]$ direction in Fig.~\ref{fig:2}(d). A more detailed formal analysis on the connection between the orbital toroidization density~$\textbf{t}(\textbf{k})$ and topological nodal lines is presented in App.~\ref{app:nodal}. 

\begin{figure*}[t!]
\centering
  \includegraphics[width=\linewidth]{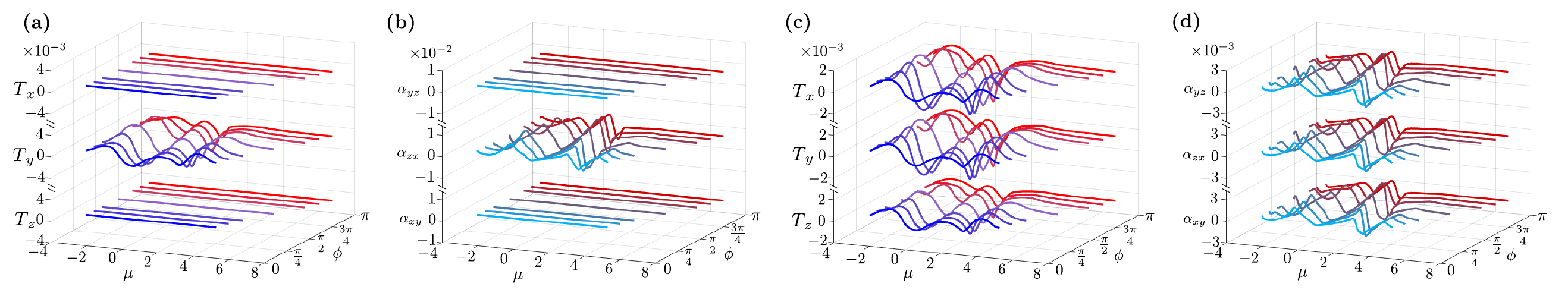}
  \caption{Controlling frustrated ferrotoroidic orbital magnetic orders. {\bf (a)} Orbital toroidization $(T_{x},T_{y},T_{z})$ and {\bf (b)} asymmetric magnetoelectric tensor $(\alpha_{xy},\alpha_{yz},\alpha_{zx})$ components as a function of chemical potential $\mu$ and magnetic flux $\phi$, in a 2-in-2-out pattern. 
  {\bf (c)}, {\bf (d)} Same for a 3-in-1-out pattern. The anisotropic magnetoelectric tensor components can be experimentally accessed via zero-field current-induced Hall effects.}
\label{fig:3}
\end{figure*}

Further to the momentum-space resolution of the orbital toroidization components $T_k$, we also show how the net toroidization $\textbf{T}$ changes with chemical potential $\mu$ and magnetic flux associated with the Aharonov--Bohm phase $\phi$, at~$T=0$~K, in Fig.~\ref{fig:3}(a) and Fig.~\ref{fig:3}(c). As expected from symmetry considerations, toroidization develops exclusively in one Cartesian coordinate for a 2-in-2-out configuration, whereas it has equal magnitude in all three components for a 1-in-3-out configuration. This is consistent with the six \mbox{2-in-2-out} orders corresponding to the six $\langle 100 \rangle$ \mbox{vectors}, and~with the eight 1-in-3-out and 3-in-1-out orders corresponding to the eight $\langle 111 \rangle$ vectors. The sign and the magnitude of the toroidization vary with filling, and are thus controlled by the chemical potential $\mu$. The toroidization changes periodically with $\phi$, satisfying ${\mathbf{T}(\phi)=-\mathbf{T}(2\pi-\phi)}$ and ${\mathbf{T}(\phi)=\mathbf{T}(\phi+2\pi)}$. \mbox{Notably}, the largest toroidization magnitude~$\vert \mathbf{T} \vert$ occurs between $\phi =\frac{\pi}{3}$ and $\phi =\frac{2\pi}{3}$. 

We now focus on the physical manifestations of orbital toroidization in FTOMs. In low-energy electromagnetic \mbox{responses}, the presence of ground-state orbital toroidization $\textbf{T}$ introduces an asymmetry in the magnetoelectric tensor $\alpha_{ij} \neq \alpha_{ji}$, for ${i \neq j}$~\cite{Dubovik1990,Spaldin2008}. The magnetoelectric tensor, $\alpha_{ij}$, is related to the magnetic quadrupole moment, $\mathcal{Q}_{ij}$, through $\alpha_{ij} = -\frac{\partial \mathcal{Q}_{ij}}{\partial \mu}$~\cite{GaoQ}. Therefore, for an antisymmetric~$\mathcal{Q}_{ij}$, the magnetoelectric tensor can be directly related to toroidization: $\alpha_{ij} = - \varepsilon_{ijk} \frac{\partial {T}_{k}}{\partial \mu}$~\cite{Gao2018Spin}. We plot three of the components of $\alpha_{ij}$ as a function of chemical potential $\mu$ and flux $\phi$ at zero temperature in Fig.~\ref{fig:3}(b) and Fig.~\ref{fig:3}(d). We note that in our $\mathcal{PT}$ symmetric models, the diagonal components, $\alpha_{ii}$, vanish. The remaining three components are related to those plotted by $\alpha_{ij} = -\alpha_{ji}$. 

As a direct physical consequence of their magnetoelectric properties, the frustrated toroidal magnets can realize a zero-field current-induced Hall effect~\cite{Khomskii2014, Mineev2025}. Namely, a \mbox{current} $\textbf{j}$ flowing through a FTOM leads to a magnetization $\textbf{M}$ that develops in a direction perpendicular to the current ($\textbf{M} \perp \textbf{j}$) and toroidization ($\textbf{M} \perp \textbf{T}$). The orbital magnetization $\textbf{M}$ gives rise to a~Hall-like response with the induced transverse electric field $E_j$ proportional to the current squared, $E_{j} = \frac{1}{nec} \alpha_{ji} \rho_{ii} j^{2}_{i}$, where $n$ and $e$ are the charge carrier density and elementary charge, $\rho_{ii}$ denotes the longitudinal resistivity, and $j_{i}$ is the applied current component. Thus, a combined observation of the current-induced zero-field Hall phenomena with the measurement of standard longitudinal resistivity allows one to reconstruct the magnetoelectric tensor components $\alpha_{ij}$ shown in Fig.~\ref{fig:3}(b) and Fig.~\ref{fig:3}(d), and to experimentally identify the frustrated toroidal magnetic order. 
%
%

\section{Frustrated topological Weyl magnetism}
We next consider the effective models of FTOMs with ground state orbital magnetization~\textbf{M}. Here, the staggered fluxes on the kagome planes mirror those of the Haldane model~\cite{Haldane1988}. The corresponding systems break $\mathcal{T}$-symmetry, but preserve $\mathcal{P}$-symmetry.

We find that such frustrated orbital magnets exhibit Weyl nodes, i.e., linearly dispersing band touchings which act as sources and sinks of Berry curvature, ${\boldsymbol\Omega^n = \text{i} \boldsymbol\nabla_\textbf{k} \times \textbf{A}_{nn}}$~\cite{Armitage2018}. In pyrochlore FTOM models, the Weyl nodes arise between bands $n=2,3$, see App.~\ref{app:models}. The Weyl nodes are characterized by two parameters: $\mathbf{b}$, the displacement vector between the two nodes in $\textbf{k}$-space, and $b_{0}$, the \mbox{energy} difference between the nodes~\cite{Vazifeh2013}. The direction of $\mathbf{b}$ is related to the symmetries of the system in an analogous way to the toroidization, and $b_{0}$ vanishes due to $\mathcal{P}$-symmetry. We comment on possible $\mathcal{P}$-symmetry breaking effects in App.~\ref{app:Psym}. 

\begin{figure*}[t!]
\centering
\includegraphics[width=\linewidth]{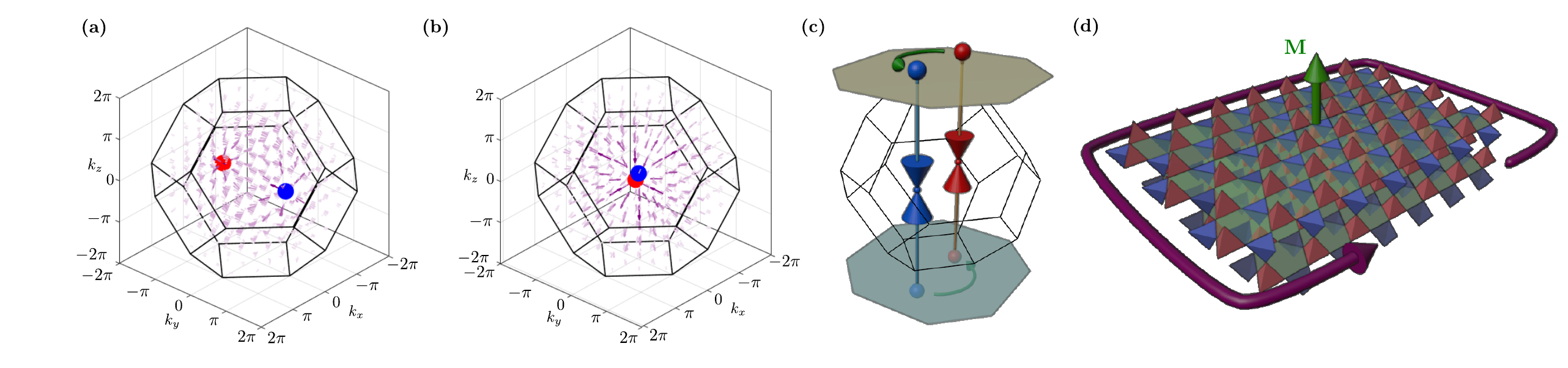}
  \caption{Frustrated orbital Weyl magnets. Topological Weyl nodes carrying positive (\text{red}) and negative (\text{blue}) Berry curvatures that arise in~\mbox{2-in-2-out} {\bf (a)} and 3-in-1-out {\bf (b)} frustrated orbital magnetic orders. {\bf (c)} Fermi arcs arising in frustrated orbital Weyl magnets on the \mbox{pyrochlore} lattice. {\bf (d)} Bulk orbital magnetization and chiral edge current due to the Weyl nodes in a~pyrochlore FTOM.}
\label{fig:4}
\end{figure*}

In Fig.~\ref{fig:4}, we highlight the role of the Weyl nodes as sources and sinks of Berry curvature in pyrochlore FTOMs, showing how they allow to discern distinct frustrated flux configurations. Fig.~\ref{fig:4}(a) and Fig.~\ref{fig:4}(b) demonstrate the Berry curvature vector $\boldsymbol\Omega_n$ throughout the BZ for both the \mbox{2-in-2-out} and \mbox{1-in-3-out} FTOM configurations, when the chemical potential $\mu$ matches the energy of the Weyl nodes.

The Weyl nodes separate the BZ into regions with two-dimensional slices of different Chern number~\cite{Armitage2018}. Slices with nonzero Chern number then contribute to the existence of edge states forming a surface band, where the constant energy contour at the Fermi energy is referred to as a Fermi arc, shown in Fig.~\ref{fig:4}(c). This leads to a Hall conductance in the plane with normal vector parallel to $\mathbf{b}$, and with magnitude proportional to the length of this vector. The value of $|\mathbf{b}|$ changes with the staggered magnetic flux $\phi$ in the FTOMs.

Additionally, 1-in-3-out and 3-in-1-out configurations are gapped between the third and fourth bands and have nonzero Chern number for all slices across the BZ. We identify these configurations as stacked, i.e., layered, Chern insulators~\cite{Halperin1987,Burkov2011}, when the chemical potential~$\mu$ lies in the gap. As such, the system has chiral edge states with quantized Hall conductance, i.e., it exhibits the quantized anomalous Hall effect, see Fig.~\ref{fig:4}(d). Under a suitable $\mathcal{C}_{2}$-symmetry-breaking (\mbox{$\mathcal{C}_{3}$-symmetry-breaking}) perturbation for a \mbox{2-in-2-out} (\mbox{1-in-3-out}) configuration, a gap can be opened between both bands $n=1,2$, and $n=3,4$. This leads to the formation of Chern insulator phases with opposite Hall \mbox{conductances} at the corresponding gap fillings, while \mbox{preserving} the Weyl nodes between the bands $n=2,3$.

The relevant ground-state magnetic order arising in this class of FTOMs is the net bulk orbital magnetization~$\textbf{M}$, which can be computed from the eigenstates $\ket{u_{n\mathbf{k}}}$~\cite{Ceresoli2006,Resta2006},
\begin{equation}
\label{eq:1}
{M}_{i}=\frac{1}{2} \varepsilon_{ijl} \Im\left[ \sum_{n} \int_\text{BZ} \frac{\text{d}\mathbf{k}}{(2\pi)^3}  f_{n\mathbf{k}} \langle \partial_{j} u_{n\mathbf{k}} | \mathcal{H} + \varepsilon_{n\mathbf{k}} - 2 \mu | \partial_{l} u_{n\mathbf{k}}\rangle 
\ \right],
\end{equation}
where $\partial_{j} = [\nabla_\textbf{k}]_j$, while $\varepsilon_{n\mathbf{k}}$ and $f_{n\mathbf{k}}$ are the single-particle band energies and occupations, respectively. $\Im\left[\ldots\right]$ denotes the imaginary part. 

\begin{figure*}[t!]
\centering
  \includegraphics[width=\linewidth]{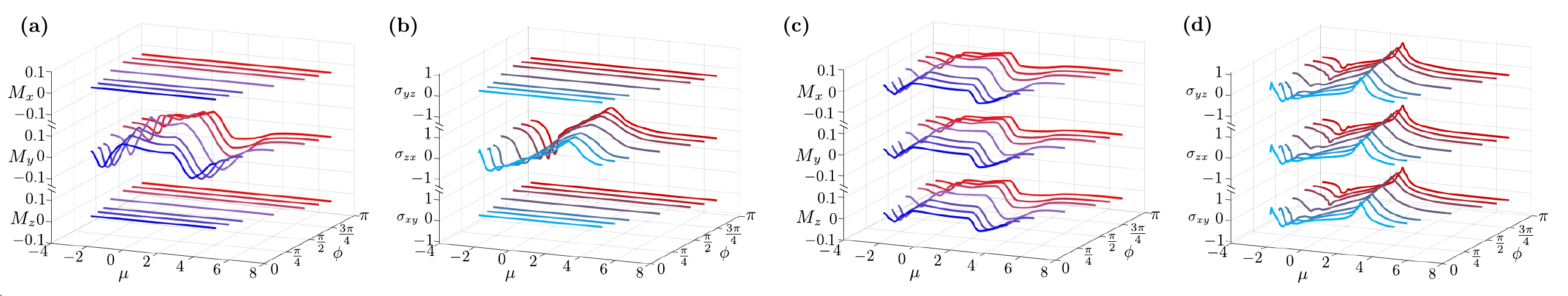}
  \caption{Probing frustrated orbital magnets with Weyl topology. {\bf (a)} Orbital magnetization $({M}_x,{M}_y,{M}_z)$ 
  and 
  {\bf (b)} anisotropic Hall conductivity tensor components $(\sigma_{xy},\sigma_{yz},\sigma_{zx})$ as a function of the chemical potential $\mu$ and flux $\phi$, for the 2-in-2-out flux configuration. 
  {\bf (c)} and {\bf (d)} same for the 3-in-1-out flux configuration. 
  }
\label{fig:5}
\end{figure*}

In Fig.~\ref{fig:5}(a) and Fig.~\ref{fig:5}(c), we plot the orbital magnetization components as a function of chemical potential~$\mu$ and flux~$\phi$ for both 2-in-2-out and 1-in-3-out configurations. Analogously to the toroidization~$\textbf{T}$, the magnitude and sign of the magnetization~$\textbf{M}$ change with chemical potential $\mu$, and their extrema vary as a function of magnetic flux. 

In Fig.~\ref{fig:5}(b) and Fig.~\ref{fig:5}(d), we further show the anomalous Hall conductivity $\sigma_{ij}=-\sigma_{ji}$ which yields the Hall currents as~$j_i =\sigma_{ij} E_j$, in 2-in-2-out and 3-in-1-out configurations. The peaks at $\mu=2$ for the 1-in-3-out configuration correspond to the Hall conductivity ${\sigma_{ij}=e^2/h}$, as expected for a Chern insulator with a Chern number $C=1$. The~peaks between $\mu=-2.5$ and $\mu=-2.0$, and those for the 2-in-2-out onfiguration, approach ${\sigma_{ij}=e^2/h}$ and become quantized upon opening a gap between the corresponding bands. The Hall conductance drops below ${\sigma_{ij}=-e^2/h}$ for a range of values of~$\phi$ due to the contributions to the Berry curvature from both band $n=1$ and the Weyl nodes between bands $n=2,3$. As~a~\mbox{consequence}, the width and the height of the low filling peaks are functions of the energy difference between the Weyl nodes and of the nodal lines between bands $n=1,2$, both of which change with the magnetic flux $\phi$.
%
%

\section{Discussion}
We now discuss scope and applicability of our findings. We~also suggest candidate material platforms that may realize the magnetic phases uncovered in our work, and we elaborate on their possible experimental validation. 

First, we argue that frustrated orbital magnetic \mbox{orders} may arise in pyrochlore iridates and ruthenates A$_2$B$_2$O$_7$ (\mbox{$\text{B}=\text{Ir, Ru}$}). The A- and B-sites form interpenetrating \mbox{pyrochlore} lattices. The A-sites host localized $f$-electron \mbox{magnetic} moments which interact with the B-site $d$-electrons. We~speculate that this interaction, at least in some limiting cases, might introduce an appropriate combination of~Aharonov--Bohm fluxes that induces frustrated orbital magnetic orders in the material. Upon leveraging such~pyrochlore systems~\cite{ZexunPriv}, e.g., Dy$_2$Ru$_2$O$_7$, FTOM phases might be obtainable under experimentally realistic conditions. Furthermore, the $f$-electron magnetism can be controlled by external magnetic fields which therefore rearrange the Aharonov--Bohm flux configurations, allowing the possibility of transitions \mbox{between} 3-in-1-out and 2-in-2-out configurations.

The distinction of 2-in-2-out and 3-in-1-out orbital \mbox{ferromagnetic} and ferrotoroidic orders relies on Hall transport measurements. Ferrotoroidic orders supporting a current-induced Hall \mbox{response} are manifested by a fundamentally different scaling of the current with electric field~(${j_i^2 \sim E_j}$), which allows to distinguish them from the ferromagnetic cases~(${j_i \sim E_j}$). The distinction between \mbox{2-in-2-out} and~\mbox{3-in-1-out} order follows from the anisotropy of the \mbox{response}, when three components of the conductivity tensor, upon gating the material in the $xy$, $yz$, and $zx$ planes, are measured. The combination of the current dependence on the gate voltage, chemical doping, and magnetic fluxes, together with the anisotropy of the response, as shown in Fig.~\ref{fig:3} and Fig.~\ref{fig:5}, allows one to unambigously and explicitly discern \mbox{distinct} frustrated topological orbital magnetic phases.

We note that only translationally symmetric systems were considered in this study. The effects of disorder, or of local interactions, deserve separate investigation. The effects of strain may also alter the measurement of the Hall responses, which invites further comment. If the strain breaks inversion symmetry in the frustrated Weyl case, nonlinear Hall effects from Berry curvature dipoles may arise~\cite{Sodemann2015}. In fact, the strain itself, acting as a~$\mathcal{P}$-breaking perturbation, can be leveraged to control the magnitude of the second-order nonlinear rectified currents, as~we show in~App.~\ref{app:Psym}. In the context of distinguishing from the ferrotoroidic case ($\textbf{T}\neq \mathbf{0}$), we stress that the possible emergence of nonlinear Hall effects may amplify the scaling discrepancy, as the Hall current $j_\text{H}$ scales as $j^2_\text{H} \propto E$ in the toroidal case, whereas the anomalous Hall effects in the frustrated orbital Weyl case scale as $j_\text{H} \propto E$ and $j_\text{H} \propto E^2$ with the electric field~$E$. It should be stressed that these scaling \mbox{dependencies} only hold in the clean crystal case, while further Hall current contributions or suppressions induced by scattering due to disorder, impurities, or phonons, may arise in \mbox{geometrically} frustrated materials with different \mbox{realistic} Fermi surfaces. 

Interestingly, a breaking of inversion symmetry may give rise to circular photogalvanic effects~\cite{deJuan2017}, beyond the measurable nonlinear Hall effects in the low-frequency limit of the applied voltage. While we elaborate on possible future directions below, all considered effects deserve further study under experimentally viable conditions.

Finally, we report and discuss the absence of orbital toroidization~$\textbf{T}$ and magnetization~$\textbf{M}$ in 4-in-0-out flux configurations. This can be understood as a result of the cancellation of the momentum-space toroidization density contributions from nodal lines forming a symmetric tetrahedron-like arrangement. Transitions from 4-in-0-out to 2-in-2-out or \mbox{1-in-3-out} phases may be induced by applied fields, which then result in the onset of Hall effects.

Physically, the absence of the ground-state orbital magnetization or orbital toroidization in the 4-in-0-out configuration can be understood as a consequence of the destructive interference of local currents $\textbf{j}(\textbf{r})$. The local currents $\textbf{j}(\textbf{r})$ arise from electron propagators on the pyrochlore lattice, under the given configuration of Aharonov--Bohm fluxes. This follows from noting that both the magnetization~$\textbf{M}$ and~toroidization~$\textbf{T}$ can be expressed in terms of the band-projected electronic Green's functions~\cite{ChenLee2011}, consistent with the underlying projector formulation of the \mbox{orbital} magnetization~$\textbf{M}$ and~toroidization~$\textbf{T}$, see App.~\ref{app:projector}. The~absence of orbital magnetization~$\textbf{M}$ and~toroidization~$\textbf{T}$ is \mbox{accompanied} by an exact vanishing of the local currents~$\textbf{j}(\textbf{r})$ and loop current circulations. Thus, \mbox{under} these circumstances, no other hidden magnetic order can be present in \mbox{a~4-in-0-out} configuration.
%
%

\section{Outlook}
The uncovering of distinct classes of geometrically \mbox{frustrated} topological orbital magnets invites several further pursuits on both theoretical and experimental grounds.  

Features of the pyrochlore lattice are central to the models and results presented in this work. It will be interesting to explore whether similar phenomena may be realizable in \mbox{different} lattices. FTOMs with polytope combinations other than tetrahedra may be possible, e.g., 4-in-2-out phases in frustrated \mbox{cubic} systems. 

Motivated by the analogy with spin ice physics drawn in our work by the configurations of orbital currents, it is \mbox{interesting} to speculate what might happen if the latter were governed by interactions between them, rather than being a~fixed background. One can envisage the orbital currents playing the role of spins in spin-ice-like systems, classical or quantum, with their characteristic emergent $\mathsf{U}(1)$ gauge symmetry and fractionalized excitations, and the ability to tune \mbox{between} \mbox{4-in-0-out}, 2-in-2-out and 1-in-3-out phases by suitable \mbox{applied} fields. Understanding the fate of frustrated orbital magnetism in presence of such dynamical orbital current configurations is likely a very tall order but also an exciting and promisingly rich endeavour. 

In this work, we focused on the charge transport reconstruction of geometrically frustrated magnetic loop current \mbox{orders} using Hall effects. It may be alternatively possible to \mbox{uncover} the proposed phases with optical phenomena. In particular, dichroism of circularly polarized light, as well as polar Kerr effect, may be used to determine the ground state orbital \mbox{magnetic} moments.

Finally, frustrated topological orbital magnetic phases \mbox{beyond} the frustrated toroidal and orbital magnetic Weyl phases may be possible. In particular, one could envisage magnetic orbital toroidal monopole orders, ${T_0 \propto \sum_\alpha \textbf{r}_\alpha \cdot \textbf{T}^\text{loc}_\alpha}$~\cite{Yamanaka2026}, with local orbital toroidizations $\textbf{T}^\text{loc}_\alpha$ at~\mbox{positions} $\textbf{r}_\alpha$. If such \mbox{geometrically} frustrated orbital orders are present, we expect higher-gradient Hall effects to emerge~\cite{Yamanaka2026}, where the ordinary anomalous Hall responses to uniform electric fields vanish, and the system responds only to nonuniform electric fields.

To sum up, in this work, we uncovered distinct classes of frustrated topological orbital magnetic orders underpinned by local loop current configurations in \mbox{orbital} ferromagnetic and ferrotoroidic ground states with \mbox{nontrivial} band topology. Our~results pave the way for a diverse range of theoretical and experimental studies, ranging from the generalizations and \mbox{extended} classifications of these phases, field-theoretic \mbox{studies}, to the design of new materials and \mbox{experimental} \mbox{protocols} that may probe these unusual \mbox{magnetic~orders}.
%
%

\section*{Acknowledgments} 
We thank Ze-Xun Lin for useful discussions. 
W.J.J.~acknowledges funding from the Rod Smallwood Studentship at Trinity College, Cambridge. 
This work was supported in part by the Engineering and Physical Sciences Research Council (EPSRC) grant No.~EP/V062654/1 (C.C.). 
%
%
%
\appendix

\begin{figure*}[t!]
\centering
\includegraphics[width=\linewidth]{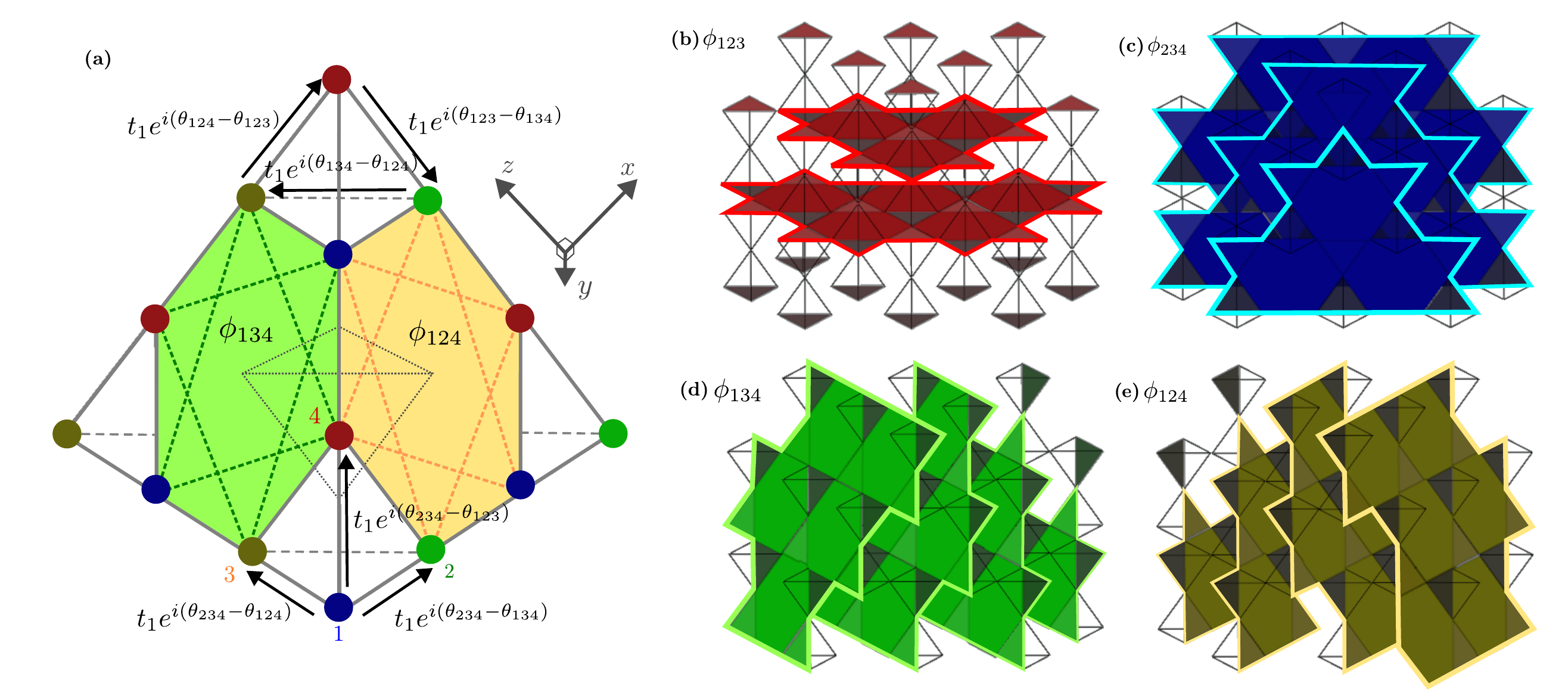}

  \caption{{\bf (a)} Portion of a~pyrochlore lattice highlighting the position of the lattice sites, labeled as $i=1-4$ in different colors. Two of the hexagons belonging to different kagome planes, through which the staggered fluxes are threaded, and their corresponding $\phi_{ijk}$, are shown. The Aharonov--Bohm phase factors $\theta_{ijk}$ acting on the nearest neighbor hoppings can be used to break $\mathcal{P}$-symmetry in the chiral pattern, with the corresponding hoppings in the opposite direction picking up an opposite phase. {\bf (b)}--{\bf(d)} Kagome planes of the pyrochlore lattice and the corresponding fluxes $\phi_{ijk}$. The~color of each set of kagome planes is chosen such that it matches that of the site not included in it in panel {\bf (a)}.}
\label{fig:6}
\end{figure*}

\section{\label{app:models} Further details on pyrochlore lattice and models}
Here we provide further details on the pyrochlore lattice and tight-binding models of FTOMs. The pyrochlore lattice is a regular arrangement of corner-sharing tetrahedra, illustrated in Fig.~\ref{fig:6}. The centres of the tetrahedra form a diamond lattice, which is bipartite and distinguishes between `up-' and `down-tetrahedra'. In order to label the pyrochlore lattice sites, one can conveniently address only up-tetrahedra and their four corner sites. From this construction one can see that the pyrochlore lattice is tetrapartite, and that its primitive unit cells form a face-centred cubic (fcc) lattice --- namely, the lattice formed by the up-tetrahedra, equivalent to one sublattice of the diamond lattice. In the conventional cubic unit cell, the vectors connecting the center of a tetrahedron to its corners are parallel to the local $[111]$ direction. Four up-tetrahedra are then connected via their nearest corners pairwise by the down-tetrahedra, see Fig.~\ref{fig:6}. 

It is important for our work to note that the pyrochlore lattice can also be viewed along one of the global $[111]$ axes as a~layered arrangement of alternating kagome and triangular lattices. Given one of the four possible choices of global $[111]$ axis, the corresponding triangular layer contains all sites belonging to one sublattice (e.g.,~$i=1$), whereas the kagome planes contain all other sublattices (e.g.,~\mbox{$i=2,3,4$}). 

One can also observe that sets of four closest up-tetrahedra on the pyrochlore lattice form super-tetrahedra, illustrated by the four white tetrahedra in Fig.~\ref{fig:6}(a). The four faces of each super tetrahedron contain one hexagon of a kagome plane of corresponding orientation, indeed for a total of four distinct kagome plane orientations, also shown in Figs.~\ref{fig:6}(b)--(d). 
The centers of the hexagons form in turn a pyrochlore lattice, which is identical to the original pyrochlore lattice up to some displacement vector. This can be seen in Fig.~\ref{fig:6}(a) — and also Fig.~\ref{fig:1} — where connecting the centers of the four hexagons in all possible ways forms a down-tetrahedron, sitting precisely at the center of the super-tetrahedron. It is with respect to this dual pyrochlore lattice that the patterns of orbital magnetization currents are labeled \mbox{3-in-1-out}, 2-in-2-out, and 4-in-0-out in our work. 

\begin{figure*}[t!]
\centering
\includegraphics[width=\linewidth]{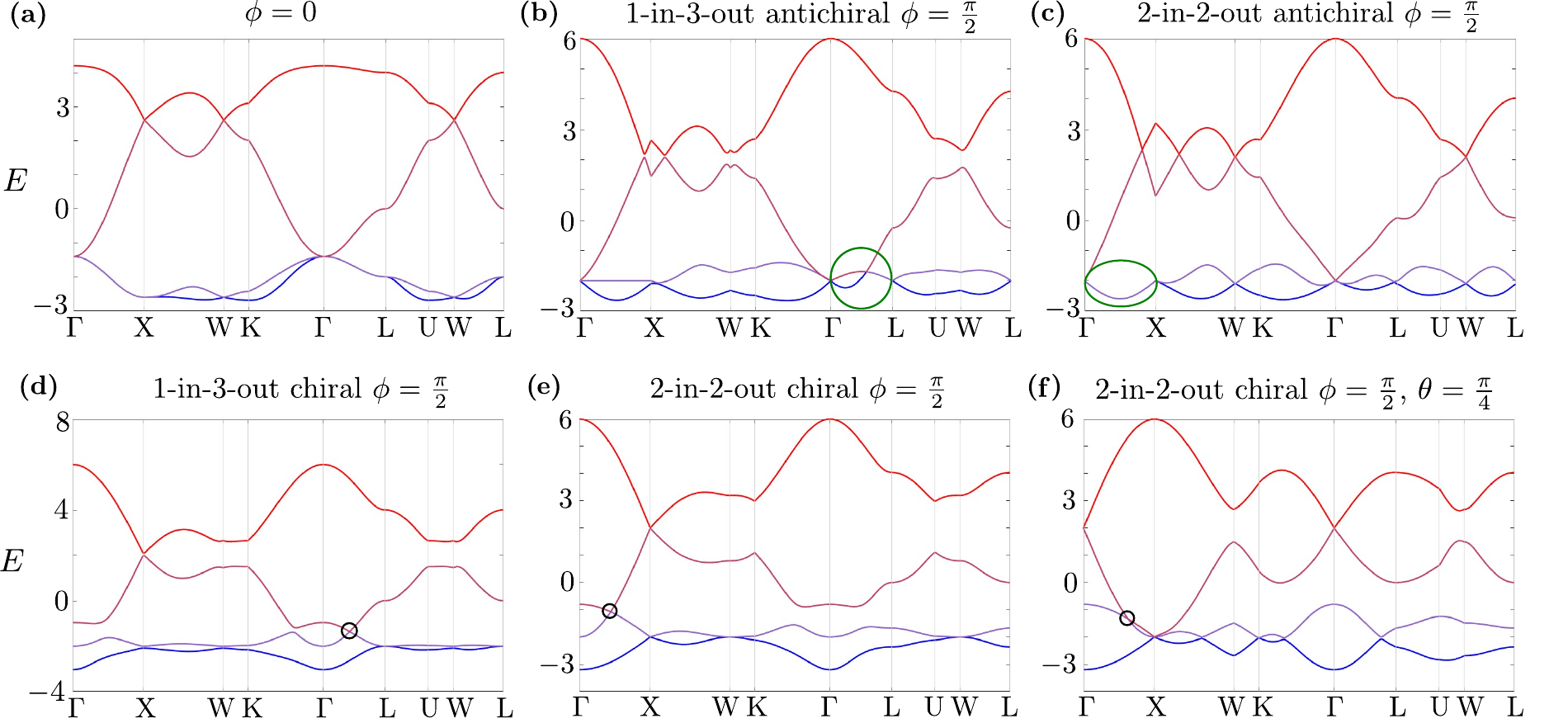}
  \caption{Band structure of the model with $t_{1}=-1$ and $t_{2}=0.15$, for different flux patterns. The Weyl points are marked by small black circles, and the topological nodal lines by green ovals. {\bf (a)} No flux threading any of the kagome planes. {\bf (b)},~{\bf (c)} Staggered fluxes with antichiral pattern of Aharonov--Bohm phases for 1-in-3-out and 2-in-2-out configurations, respectively. {\bf (d)},~{\bf (e)} Staggered fluxes with chiral pattern of Aharonov--Bohm phases for 1-in-3-out and 2-in-2-out configurations, respectively. {\bf (f)} Staggered fluxes with chiral pattern of Aharonov--Bohm phases for a 2-in-2-out configuration and the corresponding $\mathcal{P}$-symmetry breaking perturbation preserving $\mathcal{C}_{4}$-symmetry.}
\label{fig:7}
\end{figure*}

The BZ is that of an fcc lattice, namely a truncated octahedron, shown in Fig.~\ref{fig:2} and Fig.~\ref{fig:4}. We label the customary high symmetry points at the \mbox{crystal} momenta $\textbf{k}=(k_x, k_y,k_z)$ appearing in Fig.~\ref{fig:7} by: ${\Gamma = (0,0,0)}$, ${\text{X}= (0,2\pi,0)}$, ${\text{W}=(2\pi,\pi,0)}$, ${\text{K}=(\frac{3\pi}{4},\frac{3\pi}{4},0)}$, ${\text{L} =(\pi,\pi,\pi)}$, and ${\text{U}=(2\pi,\frac{\pi}{2},\frac{\pi}{2})}$. 

Let ${i,j = 1, 2, 3, 4}$ denote the four sites on an up-tetrahedron of the pyrochlore lattice with positions $ \mathbf{r}_1 = \frac{1}{4} (0,0,0)$, \mbox{$\mathbf{r}_2 = \frac{1}{4} (1,1,0)$}, \mbox{$\mathbf{r}_3 = \frac{1}{4} (0,1,1)$} and $ \mathbf{r}_4 = \frac{1}{4} (1,0,1)$, which \mbox{defines} our origin and choice of global reference frame. We introduce the displacement vectors $\mathbf{r}_{ij} = \mathbf{r}_{j} - \mathbf{r}_{i}$. The tight-binding Hamiltonians $H$ then reads 
\beq{}
    H = \sum_{i,j} \hat{c}^\dagger_{i\textbf{k}} \mathcal{H}_{ij}(\textbf{k}) \hat{c}_{j\textbf{k}},
\eeq
with creation and annihilation operators $\hat{c}^\dagger_{i\textbf{k}}$ and $\hat{c}_{i\textbf{k}}$ for electrons in the Bloch wave orbitals ${\phi_{i\textbf{k}}(\textbf{r}) = \frac{1}{\mathcal{N}} \sum_{l} e^{i\textbf{k} \cdot(\textbf{r}_i +\textbf{R}_l)} \phi_i(\textbf{r}-{\textbf{r}_i-\textbf{R}_l})}$. The Bloch \mbox{orbitals} are constructed from local orbitals $\phi_i(\textbf{r})$ centered at sublattice position~$\textbf{r}_i$, with lattice vectors $\textbf{R}_l$, and index $l$ labeling unit cells. $\mathcal{N}$ is an overall normalization factor. The Bloch Hamiltonian $\mathcal{H}_{ij}(\textbf{k})$ includes nearest neighbor (NN) and next-nearest neighbor (NNN) hopping terms,
\beq{}
    \mathcal{H}_{ij}(\textbf{k}) = \mathcal{H}_{ij}^{\text{NN}}(\textbf{k}) + \mathcal{H}_{ij}^{\text{NNN}}(\textbf{k})
    \, . 
\eeq
The nearest neighbor (NN) hopping Hamiltonian reads
\begin{equation}
\label{eq:1}
\mathcal{H}_{ij}^{\text{NN}}(\textbf{k})= - 2 t_{1} \cos (\mathbf{k} \cdot \mathbf{r}_{ij} ),  \qquad \langle ij\rangle,
\end{equation}
where $\langle i j \rangle$ denotes NN sites. We further include NNN hoppings on the kagome planes of all pyrochlore lattice, see~Fig.~\ref{fig:6}. Two different patterns of staggered fluxes threading the hexagons of the kagome planes are considered, yielding the different topological features and responses studied in~the main text.

We label the fluxes threading each of the hexagons as $\phi_{123}$, $\phi_{124}$, $\phi_{134}$ and $\phi_{234}$, where the triplet of indices denotes the pyrochlore sites involved in the corresponding kagome plane direction, see Fig.~\ref{fig:6}. Due to the periodicity with $\phi_{ijk}$ of the following models, we choose ${\phi_{ijk} \in [-\pi, \pi)}$, with \mbox{positive} values corresponding to `out' and negative values to `in'. The `all out' configuration has ${\phi_{123} = \phi_{124} = \phi_{134} = \phi_{234}}$, whereas a 2-in-2-out configuration has ${(\phi_{123},\phi_{124},\phi_{134}, \phi_{234})= (\phi, \phi, -\phi, -\phi)}$, with positive $\phi$.

The first (antichiral) pattern of fluxes explored in the main text, which yields ground state toroidal order, is shown in Fig.~\ref{fig:1}(b). For~this model, the NNN Bloch Hamiltonian \mbox{matrix} elements read:
\beq{}
\begin{aligned}
\mathcal{H}_{12}^{\text{NNN}}(\mathbf{k}) = &- 2 t_{2} [ \cos (\mathbf{k} \cdot (\mathbf{r}_{13} + \mathbf{r}_{23}) + \phi_{123}) 
\\
&+ \cos (\mathbf{k} \cdot (\mathbf{r}_{14} + \mathbf{r}_{24}) + \phi_{124})], \\
\mathcal{H}_{13}^{\text{NNN}}(\mathbf{k})= &- 2 t_{2} [ \cos (\mathbf{k} \cdot (\mathbf{r}_{12} - \mathbf{r}_{23}) + \phi_{123})\\
&+ \cos (\mathbf{k} \cdot (\mathbf{r}_{14} + \mathbf{r}_{34}) + \phi_{134})], \\
\mathcal{H}_{14}^{\text{NNN}}(\mathbf{k}) = &- 2 t_{2} [ \cos (\mathbf{k} \cdot (\mathbf{r}_{12} - \mathbf{r}_{24}) + \phi_{124})\\
&+ \cos (\mathbf{k} \cdot (\mathbf{r}_{13} - \mathbf{r}_{34}) + \phi_{134})], \\
\mathcal{H}_{23}^{\text{NNN}}(\mathbf{k})= &- 2 t_{2} [ \cos (\mathbf{k} \cdot (-\mathbf{r}_{12} - \mathbf{r}_{13}) + \phi_{123})\\
&+ \cos (\mathbf{k} \cdot (\mathbf{r}_{24} + \mathbf{r}_{34}) + \phi_{234})], \\
\mathcal{H}_{24}^{\text{NNN}}(\mathbf{k})= &- 2 t_{2} [ \cos (\mathbf{k} \cdot (-\mathbf{r}_{12} - \mathbf{r}_{14}) + \phi_{124})\\
&+ \cos (\mathbf{k} \cdot (\mathbf{r}_{23} - \mathbf{r}_{34}) + \phi_{234})], \\
\mathcal{H}_{34}^{\text{NNN}}(\mathbf{k})= &- 2 t_{2} [ \cos (\mathbf{k} \cdot (-\mathbf{r}_{13} - \mathbf{r}_{14}) + \phi_{134})\\
&+ \cos (\mathbf{k} \cdot (-\mathbf{r}_{23} - \mathbf{r}_{24}) + \phi_{234})],  \\
\end{aligned}
\eeq
This model preserves $\mathcal{PT}$-symmetry, as can be inferred from all the elements of the Hamiltonian being real~\cite{Wu2019}, while the $\mathcal{P}$- (inversion) and $\mathcal{T}$-symmetry ({time-reversal}) are \mbox{broken} \mbox{independently}. This combination of symmetries is of particular interest, as it both allows for nodal lines to be assigned a~topological charge and supports toroidal order. These topological nodal lines can be seen in Fig.~\ref{fig:7}(b) and Fig.~\ref{fig:7}(c) for both 1-in-3-out and 2-in-2-out configurations \mbox{between} $\Gamma$~and~L, or~$\Gamma$~and~X points, \mbox{respectively}.

The second (chiral) pattern of fluxes is shown in Fig.~\ref{fig:1}(d), for which the elements of the NNN Bloch Hamiltonian read:

\beq{}
\begin{aligned}
\mathcal{H}_{12}^{\text{NNN}}(\mathbf{k})= &- 2 t_{2} [ e^{i\phi_{123}} \cos (\mathbf{k} \cdot (\mathbf{r}_{13} + \mathbf{r}_{23})) 
\\
&+ e^{-i\phi_{124}}\cos (\mathbf{k} \cdot (\mathbf{r}_{14} + \mathbf{r}_{24}))], \\
\mathcal{H}_{13}^{\text{NNN}}(\mathbf{k})= &- 2 t_{2} [ e^{-i\phi_{123}}\cos (\mathbf{k} \cdot (\mathbf{r}_{12} - \mathbf{r}_{23}))\\
&+ e^{i\phi_{134}}\cos (\mathbf{k} \cdot (\mathbf{r}_{14} + \mathbf{r}_{34}))], \\
\mathcal{H}_{14}^{\text{NNN}}(\mathbf{k})= &- 2 t_{2} [ e^{i\phi_{124}}\cos (\mathbf{k} \cdot (\mathbf{r}_{12} - \mathbf{r}_{24}))\\
&+ e^{-i\phi_{134}}\cos (\mathbf{k} \cdot (\mathbf{r}_{13} - \mathbf{r}_{34}))], \\
\mathcal{H}_{23}^{\text{NNN}}(\mathbf{k})= &- 2 t_{2} [ e^{i\phi_{123}}\cos (\mathbf{k} \cdot (-\mathbf{r}_{12} - \mathbf{r}_{13}))\\
&+ e^{-i\phi_{234}}\cos (\mathbf{k} \cdot (\mathbf{r}_{24} + \mathbf{r}_{34}))], \\
\mathcal{H}_{24}^{\text{NNN}}(\mathbf{k})= &- 2 t_{2} [ e^{-i\phi_{124}}\cos (\mathbf{k} \cdot (-\mathbf{r}_{12} - \mathbf{r}_{14}))\\
&+ e^{i\phi_{234}}\cos (\mathbf{k} \cdot (\mathbf{r}_{23} - \mathbf{r}_{34}))], \\
\mathcal{H}_{34}^{\text{NNN}}(\mathbf{k})= &- 2 t_{2} [ e^{i\phi_{134}}\cos (\mathbf{k} \cdot (-\mathbf{r}_{13} - \mathbf{r}_{14}))\\
&+ e^{-i\phi_{234}}\cos (\mathbf{k} \cdot (-\mathbf{r}_{23} - \mathbf{r}_{24}))].\\
\end{aligned}
\eeq
This model breaks $\mathcal{T}$-symmetry and preserves $\mathcal{P}$-symmetry, thus admitting nonvanishing Berry curvature. Analogously to the former, the topological feature, here the~Weyl nodes, can be seen in Fig.~\ref{fig:7}(d) and Fig.~\ref{fig:7}(e) for both 1-in-3-out and~\mbox{2-in-2-out} configurations between $\Gamma$~and~L, or~$\Gamma$~and~X, respectively, upon inspection of bands $n=2,3$.

In order to exhibit a nonlinear Hall effect related to the Weyl node dipole, a suitable $\mathcal{P}$-symmetry breaking perturbation separating the Weyl nodes in energy must be applied. We~consider a perturbation changing the nearest neighbor hoppings as~follows:
\beq{}
\begin{aligned}
\mathcal{H}_{12}^{\text{NN}}(\mathbf{k})=& - 2 t_{1}   \cos (\mathbf{k} \cdot \mathbf{r_{12}} - \theta_{134} + \theta_{234}),\\
\mathcal{H}_{13}^{\text{NN}}(\mathbf{k})=& - 2 t_{1}   \cos (\mathbf{k} \cdot \mathbf{r_{13}} - \theta_{124} + \theta_{234}),\\
\mathcal{H}_{14}^{\text{NN}}(\mathbf{k})=& - 2 t_{1}   \cos (\mathbf{k} \cdot \mathbf{r_{14}} - \theta_{123} + \theta_{234}),\\
\mathcal{H}_{23}^{\text{NN}}(\mathbf{k})=& - 2 t_{1}   \cos (\mathbf{k} \cdot \mathbf{r_{23}} - \theta_{124} + \theta_{134}),\\
\mathcal{H}_{24}^{\text{NN}}(\mathbf{k})=& - 2 t_{1}   \cos (\mathbf{k} \cdot \mathbf{r_{24}} - \theta_{123} + \theta_{134}),\\
\mathcal{H}_{34}^{\text{NN}}(\mathbf{k})=& - 2 t_{1}   \cos (\mathbf{k} \cdot \mathbf{r_{34}} - \theta_{123} + \theta_{124}).\\
\end{aligned}
\eeq
This way, if the different $\theta_{ijk}$ are chosen analogously to $\phi_{ijk}$ in a 1-in-3-out or 2-in-2-out configuration, the relevant $\mathcal{C}_{n}$-symmetry is preserved, and the location of the Weyl nodes between bands $n=2,3$ remains confined to the \mbox{corresponding} path connecting high symmetry points. This can be seen in~Fig.~\ref{fig:7}(f), where both $\phi$ and $\theta$ are chosen in a 2-in-2-out configuration, and the Weyl nodes stay on the line connecting $\Gamma$~and~X points.

For all quantities computed in the main text and App.~\ref{app:Psym}, the~hopping amplitudes were set to ${t_{1}=-1}$ and ${t_{2}=0.15}$. The order-of-magnitude lower values of $t_2 \ll t_1$ are motivated by realistic values of NNN hoppings in candidate \mbox{materials}~\cite{Witczak2013, Shinaoka2015, Zhang2017}. 

%
%

\section{\label{app:projector}
Projector formulation of orbital toroidization and magnetization}
In the following, we provide gauge-invariant formulations of orbital toroidizations and magnetizations, in terms of Bloch band projectors, $P_n = \ket{u_{n\textbf{k}}}\bra{u_{n\textbf{k}}}$. We start from the expression for the orbital magnetic quadrupole moment $\mathcal{Q}_{ij}$ of Ref.~\onlinecite{GaoQ}, which we reproduce here for convenience: 
\begin{widetext}
\begin{equation}
\begin{aligned}
\mathcal{Q}_{ij} =\sum_{n} \int_{\text{BZ}} \frac{d\mathbf{k}}{(2\pi)^3} \left\{ f_{n\mathbf{k}} \left[ \sum_{m\neq n} \frac{2}{3} \Re \{ A^{i}_{nm}M^{j}_{mn}\} -\frac{1}{12} \varepsilon_{klj} (\partial_{k}\Gamma^{n}_{il})\right] + \mathcal{G}_{n\mathbf{k}} \left[ - \sum_{m\neq n} \frac{2 \Re\{A^{i}_{nm}M^{j}_{mn}\}}{\epsilon_{n\mathbf{k}}-\epsilon_{m\mathbf{k}}} - \frac{1}{3}\varepsilon_{klj}(\partial_{k}g_{il}^{n})\right] \right\}.
\end{aligned}
\end{equation}
\end{widetext}
$\Re\{ \ldots \} $ denotes the real part, \mbox{$\Gamma^{n}_{ij} = \langle u_{n} | \partial_{i} \partial_j \mathcal{H} |u_{n}\rangle$} is a~\mbox{Hessian} matrix, \mbox{$\mathcal{G}_{n\mathbf{k}}=-k_{B}T \ln(1+\exp[(\mu-\epsilon_{n\mathbf{k}})/(k_{B}T)])$}, and \mbox{$M^{j}_{mn}=\sum_{p\neq n} \frac{1}{2} \varepsilon_{jkl} (v^{k}_{mp}+v^{k}_{nn} \delta_{mp})A^{l}_{pn}$}, with the \mbox{velocity} matrix elements, $v^k_{nm} = \bra{u_{n\textbf{k}}} \partial_k \mathcal{H} \ket{u_{m\textbf{k}}}$.
The quantum \mbox{metric}, \mbox{$g^{n}_{ij}=\sum_{m \neq n} \Re \{ A^{i}_{nm}A^{j}_{mn}\}$}, term can be integrated by parts, $\int \text{d}\mathbf{k}~ \mathcal{G}_{n\mathbf{k}} (\partial_{k} g_{il}^{n})=-\int \text{d}\mathbf{k}~ f_{n\mathbf{k}} v_{nn} g_{il}^{n} $, with $v_{nn}=\partial_i \varepsilon_{n\textbf{k}}$ denoting the intraband velocity. At zero temperature, ${\mathcal{G}_{n\mathbf{k}}=(\varepsilon_{n\textbf{k}}-\mu)\Theta(\mu-\varepsilon_{n\textbf{k}})}$, with $\Theta(\mu-\varepsilon)$ the Heaviside step function. $\mathcal{Q}_{ij}$ can then be expressed in terms of band \mbox{projectors} as:
\begin{widetext}
\begin{equation}
\begin{aligned}
\mathcal{Q}_{ij}
&=
\sum_n\int_{\text{BZ}}\frac{\mathrm{d}\mathbf{k}}{(2\pi)^3}\,
\epsilon_{jkl}
\Bigg\{
f_{n\mathbf{k}}
\Bigg[
\frac{1}{3}\Re\Bigg\{
\mathrm{Tr}\Big[
P_n(\partial_iP_n)
\Big\{(\partial_lP_n),(\partial_k\mathcal{H})\Big\}
\Big]
\Bigg\}
-\frac{1}{12}\mathrm{Tr}\Big[
(\partial_kP_n)(\partial_i\partial_l\mathcal{H})
+P_n(\partial_i\partial_l\partial_k\mathcal{H}_{\mathbf{k}})
\Big]
\\[-1pt]
\qquad
&+\frac{1}{3}\Re\Bigg\{
\mathrm{Tr}\Big[
P_n(\partial_k\mathcal{H})
(\partial_iP_n)(\partial_lP_n)
\Big]
\Bigg\}
\Bigg]
+\mathcal{G}_{n\mathbf{k}}
\Re\Bigg\{
\mathrm{Tr}\Bigg[
\sum_{m\neq n}
\frac{
P_n(\partial_iP_m)
\Big\{(\partial_lP_n),(\partial_k\mathcal{H})\Big\}
}{
\epsilon_{n\mathbf{k}}-\epsilon_{m\mathbf{k}}
}
\Bigg]
\Bigg\}
\Bigg\},
\end{aligned}
\end{equation}
\end{widetext}
Hence, it follows that the components of the orbital \mbox{toroidization} $\textbf{T}$, determined by $\mathcal{Q}_{ij}$ as ${T}_{k} = \frac{1}{2} \varepsilon_{ijk} \mathcal{Q}_{ij}$, can be written as:
\begin{widetext}
\begin{equation}
\begin{aligned}
T_{k}
&=
\sum_n\int_\text{BZ}\frac{\mathrm{d}\mathbf{k}}{(2\pi)^3}\,
\left(
\delta_{k j}\delta_{il}
-\delta_{k l}\delta_{ij}
\right)
\Bigg\{
f_{n\mathbf{k}}
\Bigg[
\frac{1}{3}\Re\Bigg\{
\mathrm{Tr}\Big[
P_n(\partial_iP_n)
\Big\{(\partial_lP_n),(\partial_j\mathcal{H})\Big\}
\Big]
\Bigg\}
-\frac{1}{12}\mathrm{Tr}\Big[
(\partial_jP_n)(\partial_i\partial_l\mathcal{H})
\Big]
\\[-1pt]
\qquad
&+\frac{1}{3}\Re\Bigg\{
\mathrm{Tr}\Big[
P_n(\partial_j\mathcal{H})
(\partial_iP_n)(\partial_lP_n)
\Big]
\Bigg\}
\Bigg]
+\mathcal{G}_{n\mathbf{k}}
\Re\Bigg\{
\mathrm{Tr}\Bigg[
\sum_{m\neq n}
\frac{
P_n(\partial_iP_m)
\Big\{(\partial_lP_n),(\partial_j\mathcal{H})\Big\}
}{
\varepsilon_{n\mathbf{k}}-\varepsilon_{m\mathbf{k}}
}
\Bigg]
\Bigg\}
\Bigg\}.
\end{aligned}
\end{equation}
\end{widetext}

Note that in the presence of $\mathcal{P}$-symmetry or $\mathcal{T}$-symmetry, the projectors satisfy $P(\mathbf{k})=P(-\mathbf{k})$ or $P(\mathbf{k})=P^{*}(-\mathbf{k})$, \mbox{respectively}, and analogously for the Bloch Hamiltonian~$\mathcal{H(\textbf{k})}$. Given that every term in the expression above is real by construction, or has its real part taken, the effects of these two symmetries are equivalent. Using the simpler case of $\mathcal{P}$-symmetry, we note that $P(\mathbf{k})$ and $\mathcal{H}(\mathbf{k})$ are even functions. It follows that, since every term includes the product of an odd number (three) of derivatives of these quantities, the toroidization density resolved over \mbox{$\mathbf{k}$-space}, $t_{k}(\mathbf{k})$, with ${T}_k=\int_{\text{BZ}}\frac{\mathrm{d}\mathbf{k}}{(2\pi)^3}t_{k}(\mathbf{k})$, is an odd function. Thus, integrating this quantity over the occupied subspace yields net zero toroidization, due to the equal and \mbox{opposite} \mbox{contributions} at $\mathbf{k}$ and $-\mathbf{k}$, as we discuss in the main text.

The other ground-state property considered in the main text, i.e., the \mbox{orbital} magnetization~$\textbf{M}$, can be compactly expressed in~terms of projectors as:
\begin{widetext}
\begin{equation}
\begin{aligned}
{M}_{i} = \frac{1}{2} \varepsilon_{ijl} \Im \left[ \sum_{n} \int_\text{BZ} \frac{\text{d}\mathbf{k}}{(2\pi)^3}  f_{n\mathbf{k}} \mathrm{Tr} [P_{n}(\partial_{j} P_{n})(\mathcal{H}+\varepsilon_{n\mathbf{k}}-2\mu)(\partial_{l}P_{n})] \right],
\end{aligned}
\end{equation}
\end{widetext}
which is consistent with the formulation in terms of Bloch bands $\ket{u_{n\textbf{k}}}$ in the main text.

Note that in the presence of $\mathcal{PT}$-symmetry, as is the case with the first class of models, the integrand contributions vanish identically at every point in \mbox{$\mathbf{k}$-space}.  
This is because a real gauge can be adopted, and 
the Bloch Hamiltonian and projectors are accordingly real. Each of the terms in the expression are real then, and thus vanish upon taking the imaginary part.

It is necessary for the $\mathcal{T}$-symmetry to be broken for the magnetization $\textbf{M}$ to develop, following a similar argument to the former for the nonvanishing toroidization $\textbf{T}$. In the presence of $\mathcal{T}$-symmetry, which enforces $P(\mathbf{k})=P^{*}(-\mathbf{k})$, $\mathcal{H}(\mathbf{k})=\mathcal{H}^{*}(-\mathbf{k})$ and $\varepsilon_{n\mathbf{k}}=\varepsilon_{n,-\mathbf{k}}$, it can be shown that the contributions from $\mathbf{k}$ and $-\mathbf{k}$ are equal and opposite.
\\
%
%

\section{\label{app:nodal}
Correspondence between orbital toroidization and~topological nodal lines}
In the main text we have shown numerically that there is a correspondence between orbital toroidization~\textbf{T} and~topological nodal lines demonstrated. Here we shall establish it using formal analytical arguments. Starting again from the expression for the orbital magnetic quadrupole moment $\mathcal{Q}_{ij}$, we can rearrange it and break it down into three terms, ${\mathcal{Q}_{ij} = \mathcal{Q}^{(1)}_{ij} + \mathcal{Q}^{(2)}_{ij}  + \mathcal{Q}^{(3)}_{ij}}$, which all individually vanish upon filling all bands:

\begin{widetext}
\begin{equation}
\begin{aligned}
\mathcal{Q}_{ij}^{(1)}
&={}
\sum_{n} \varepsilon_{klj}
\int_\text{BZ} \frac{\text{d}\mathbf{k}}{(2\pi)^3}\,
f_{n\mathbf{k}}
\sum_{m \neq n}
A^{i}_{nm} A^{l}_{mn}
\Bigg[
v^{k}_{nn}
\left(
\frac{2}{3}
+
\frac{\mu-\varepsilon_{n\mathbf{k}}}
{\varepsilon_{n\mathbf{k}}-\varepsilon_{m\mathbf{k}}}
\right)
+
v^{k}_{mm}
\left(
\frac{1}{3}
+
\frac{\mu-\varepsilon_{n\mathbf{k}}}
{\varepsilon_{n\mathbf{k}}-\varepsilon_{m\mathbf{k}}}
\right)
\Bigg],
\\
\mathcal{Q}_{ij}^{(2)}
&={}
-i\sum_{n} \varepsilon_{klj}
\int_\text{BZ} \frac{\text{d}\mathbf{k}}{(2\pi)^3}\,
f_{n\mathbf{k}}
\sum_{m \neq n}
\sum_{p \neq n,m}
A^{i}_{nm} A^{k}_{mp} A^{l}_{pn}
(\varepsilon_{p\mathbf{k}}-\varepsilon_{m\mathbf{k}})
\left(
\frac{1}{3}
+
\frac{\mu-\varepsilon_{n\mathbf{k}}}
{\varepsilon_{n\mathbf{k}}-\varepsilon_{m\mathbf{k}}}
\right),
\\
\mathcal{Q}_{ij}^{(3)}
&={}-\frac{1}{12} \sum_{n} \varepsilon_{klj}
\int_\text{BZ} \frac{\text{d}\mathbf{k}}{(2\pi)^3}\,
f_{n\mathbf{k}}
\Bigg[
(\partial_{i} \partial_{l} \partial_{k} \varepsilon_{n\mathbf{k}})
-2 \partial_{k}\left\{
\Re\left[
\sum_{m\neq n}
A^{i}_{nm} A^{l}_{mn}
(\varepsilon_{n\mathbf{k}}-\varepsilon_{m\mathbf{k}})
\right]
\right\}
\Bigg].
\end{aligned}
\end{equation}

The contribution to toroidization from $\mathcal{Q}_{ij}^{(1)}$ then becomes:

\begin{equation}
\begin{aligned}
T_{k}^{(1)}
=
\sum_{n}
\left(
\delta_{k j}\delta_{il}
-
\delta_{k l}\delta_{ji}
\right)
\int_\text{BZ} \frac{\text{d}\mathbf{k}}{(2\pi)^3}\,
f_{n\mathbf{k}}
\sum_{m \neq n}
A^{i}_{nm}A_{mn}^{l}
\Bigg[
v^{j}_{nn}
\left(
\frac{2}{3}
+
\frac{\mu-\varepsilon_{n\mathbf{k}}}
{\varepsilon_{n\mathbf{k}}-\varepsilon_{m\mathbf{k}}}
\right)
+
v^{j}_{mm}
\left(
\frac{1}{3}
+
\frac{\mu-\varepsilon_{n\mathbf{k}}}
{\varepsilon_{n\mathbf{k}}-\varepsilon_{m\mathbf{k}}}
\right)
\Bigg].
\end{aligned}
\end{equation}
\end{widetext}

To demonstrate the correspondence, let us set up cylindrical polar coordinates $(r,\varphi,z)$ around the nodal line, with $r$ the distance from the nodal line, $\varphi$ the azimuthal component and $z$ the coordinate along the nodal line. Assuming rotational symmetry for simplicity, we consider $A^{\varphi}_{nm}=\frac{a}{2r}$ with $a \in \mathbb{Z}$, which yields $\phi_{nm} = a\pi$ found in the main text. At first, let us consider the case where the other two components of $A_{nm}$ vanish. The surviving terms for $T_{z}^{(1)}$ are: $A^{\varphi}_{nm} A^{\varphi}_{mn} v_{nn}^{z} (\frac{2}{3}+\frac{\mu-\varepsilon_{n\mathbf{k}}}{\varepsilon_{n\mathbf{k}}-\varepsilon_{m\mathbf{k}}})$ and $A^{\varphi}_{nm} A^{\varphi}_{mn} v_{mm}^{z} (\frac{1}{3}+\frac{\mu-\varepsilon_{n\mathbf{k}}}{\varepsilon_{n\mathbf{k}}-\varepsilon_{m\mathbf{k}}})$. Let us denote $\frac{\mu-\varepsilon_{n\mathbf{k}}}{\varepsilon_{n\mathbf{k}}-\varepsilon_{m\mathbf{k}}}$ as $\xi_{nm\mathbf{k}}$ for the following discussion. It follows that 

\begin{equation}
\begin{aligned}
T^{(1)}_{z} \propto a^{2} \int \text{d}z \frac{\text{d}r}{r} f_{n\mathbf{k}}  \left[v^{z}_{nn}\Big(\frac{2}{3}+\xi_{nm\mathbf{k}}\Big) + v^{z}_{mm}\Big(\frac{1}{3}+\xi_{nm\mathbf{k}}\Big) \right].
\end{aligned}
\end{equation}

Notably, the contribution from $\int dz\frac{2}{3}f_{n\mathbf{k}} v^{z}_{nn}$ vanishes; \mbox{however}, the~remaining terms are generally nonzero for a~system with broken $\mathcal{P}$- and $\mathcal{T}$-symmetry. Similar considerations apply to $\mathbf{T}^{(3)}_{z}$, where the surviving term is of the form $A^{\varphi}_{nm}A^{\varphi}_{mn}v_{mm}^{z}$. 

Note that at $r=0$ we have $\varepsilon_{n\textbf{k}} = \varepsilon_{m\textbf{k}}$ and, ${v^{z}_{nn} = v^{z}_{mm}}$. For any band $n$, the surviving contribution is of the form ${\propto v^{z}_{nn} \xi_{nm\mathbf{k}}}$; however, at $r=0$ band $m$ is also occupied and contributes $ \propto v^{z}_{nn} \xi_{mn\mathbf{k}}$. Then, $\xi_{nm\mathbf{k}}+\xi_{mn\mathbf{k}}=-1$, and the contribution from both bands cancels upon integrating along $z$. The integral with respect to $r$ therefore does not diverge logarithmically because the radial dependence of the integrand differs from the naive $\propto \frac{1}{r}$~scaling. 

We note that $A^{r}_{nm}$ and $A^{z}_{nm}$ are generally nonzero and, \mbox{unlike} the quantized azimuthal component, can be functions of~$z$. Under the symmetry constraints of our model, contributions from these components need not vanish, but are usually much smaller close to the nodal line than~$A^{z}_{nm}$, as $A^{r}_{nm}$ and $A^{z}_{nm}$ do not scale as $\propto \frac{1}{r}$, and therefore the corresponding integral does not take the~form $\int\frac{\text{d}r}{r}\{...\}$, but instead, $\int \text{d}r~r\{...\}$.

Finally, ${T}^{(2)}_{z}$ can arise from the combination of topological nodal lines between different bands. For a variety of systems, this term is likely negligible, as it necessitates the existence of a network of topological nodal lines between many bands that are close in momentum space. The `effect' of these contributions decays as $\propto \frac{1}{r}$ and has a nondestructive orientation relative to the other terms. However, these conditions are met in our model, with the contribution from ${T}^{(2)}_{z}$ being \mbox{comparable} to the other two terms.

\section{\label{app:Psym}
Nonlinear Hall Effect in $\mathcal{P}$-symmetry broken frustrated orbital Weyl magnets}
In this Appendix, we discuss 
the nonlinear Hall effects that emerge from Berry curvature dipoles upon breaking \mbox{inversion} \mbox{symmetry} ($\mathcal{P}$) in FTOMs with Weyl topology. 

The nonlinear Hall conductivity tensor $\sigma_{ijk}$ is given by the Berry curvature dipole $D_{jl}$~as~\cite{Sodemann2015}:
 \begin{equation}\label{eq::NLHE}
 \begin{aligned}
 \sigma_{ijk}=-\varepsilon_{ilk} \frac{e^3 \tau}{2(1+i\omega\tau)}D_{jl},\\
 D_{jl}=\sum_{n} \int \frac{\text{d}\mathbf{k}}{(2\pi)^3} f_{n\mathbf{k}}(\partial_{j} \Omega_{l}^{n}),
 \end{aligned}
 \end{equation}
where $\tau$ is the relaxation time, and the frequency is set to the dc limit, ${\om \rightarrow 0}$.

At a chemical potential $\mu$ away from the energy of the Weyl nodes, the derivative of the Berry curvature is computed by differentiating its projector expression:
\begin{equation}
 \begin{aligned}
\Omega^n_{i}=\varepsilon_{ijl} \Im \left[ \mathrm{Tr}\{P_n (\partial_{j} P_n)(\partial_{l}P_n)\}\right].
 \end{aligned}
 \end{equation}

\begin{figure*}[t!]
\centering
\includegraphics[width=\linewidth]{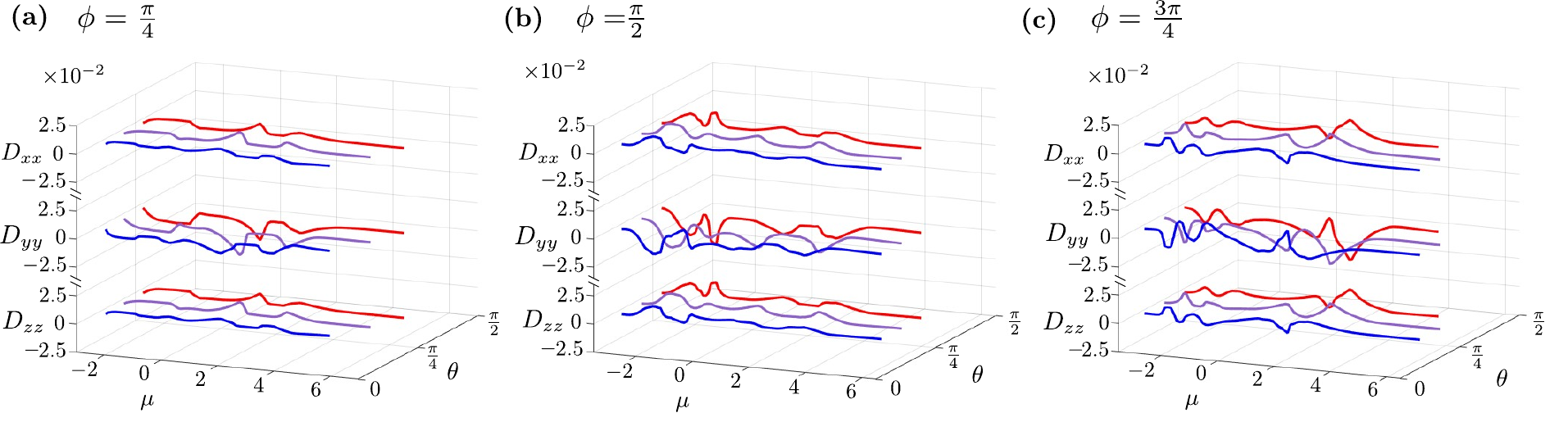}
  \caption{Nonlinear Hall effects in noncentrosymmetric frustrated orbital magnets with Weyl topology. {\bf (a)}--{\bf (c)} Diagonal components of $D_{ij}$ for~a~2-in-2-out configuration of staggered fluxes and for nearest neighbor phases, $\theta$, that break the inversion symmetry across the $y$-direction, for~{\bf (a)}~$\phi=\frac{\pi}{4}$, {\bf (b)} $\phi =\frac{\pi}{2}$ and {\bf (c)}~$\phi=\frac{3\pi}{4}$. The Berry curvature dipoles and nonlinear Hall effects are evaluated at $\theta=\frac{\pi}{12},\frac{\pi}{6}$, $\frac{\pi}{4}$.}
\label{fig:8}
\end{figure*}

To deal with the divergence of the Berry curvature at a Weyl node, an analytical approach can be used when the chemical potential is close to the Weyl node energy. The Hamiltonian, projected onto the relevant bands and close to the Weyl nodes, may be written as:
\beq{}
\mathcal{H}(\mathbf{q}) = (E_{W}+\mathbf{t} \cdot \mathbf{q}) \sigma_{0} + \sum_{i,j} v_{ij} q_{j} \sigma_{i},
\eeq
where $E_{W}$ is the energy of the Weyl node, $\mathbf{t}$ is the tilt, $v_{ij}$ are the elements of~the 3$\times$3 velocity matrix and $\sigma_{i}$ denotes the Pauli matrices. The momentum coordinate $\mathbf{q}$ is chosen such that the Weyl node lies at $\mathbf{q}=\mathbf{0}$. The topological charge of the Weyl node, i.e., chirality $\chi$, is defined by the determinant of~the velocity matrix~\cite{Armitage2018}.

We may define the matrix $G=v^{T}v$ and choose a coordinate basis $(q_{1}, q_{2}, q_{3})$ in which it is diagonal, such that $v_{i}=\sqrt{G_{ii}}$. Around the Weyl node, the energies of the corresponding bands are given by: $E(\mathbf{q})=E_{W}+\mathbf{t} \cdot \mathbf{q} \pm \sqrt{v^{2}_{1}q^{2}_{1} +v^{2}_{2}q^{2}_{2}+v
^{2}_{3}q^{2}_{3}}$. 
It follows that, when $|\mathbf{t}| < |\mathbf{v}|$, i.e., for ~type-I Weyl nodes~\cite{Soluyanov2015}, the Fermi surface is given by an ellipsoid with semiaxes that depend on the chemical potential $\mu$ and on the corresponding $t_{i}$ and $v_{i}$, and centered at some displacement from the Weyl node, as given by these same parameters. For~\mbox{type-II} Weyl nodes~\cite{Soluyanov2015}, $|\mathbf{t}| > |\mathbf{v}|$, the Fermi surface around each Weyl node takes the shape of a two-sheet hyperboloid. The Berry curvature of the lower band is given in this basis by:
\begin{equation}
\begin{aligned}
\Omega^n_{i}=\frac{\chi v_{1} v_{2} v_{3}}{2} \frac{q_{i}}{(v^{2}_{1}q^{2}_{1} +v^{2}_{2}q^{2}_{2}+v
^{2}_{3}q^{2}_{3})^{\frac{3}{2}}}.
\end{aligned}
\end{equation}
We can then directly evaluate Eq.~\eqref{eq::NLHE} to obtain: 
\begin{equation}
\begin{aligned}
D_{li}
&= \frac{2\pi\chi\delta_{il}}{3}\delta(\mathbf{q})  \\&+ \frac{\chi v_{1} v_{2} v_{3}}{2}
\left[
\frac{\delta_{il}\left(v^{2}_{1}q^{2}_{1}+v^{2}_{2}q^{2}_{2}+v^{2}_{3}q^{2}_{3}\right) -3v^{2}_{l}q_{i}q_{l}}
{\left(v^{2}_{1}q^{2}_{1}+v^{2}_{2}q^{2}_{2}+v^{2}_{3}q^{2}_{3}\right)^{\frac{5}{2}}}
\right].
\end{aligned}
\end{equation}
In the models considered in our work, $\mathbf{t}$ points along one of the ${\textbf{q}=(q_{1}, q_{2}, q_{3})}$ components, which forces off-diagonal ($i \neq l$) terms of the integral of $\partial_{l}\Omega_{i}$ to vanish, based on symmetry considerations. The topological term at the Weyl node, $\frac{2\pi\chi\delta_{il}}{3}\delta(\mathbf{q})$, does not contribute to the nonlinear conductivity given that at a certain chemical potential, the Weyl node is either unoccupied or occupied for both of the relevant bands that have opposite charge.

Given that the integral of $D_{11}$ over all of $\mathbf{k}$-space vanishes, one obtains,
\begin{equation}
\begin{aligned}
D_{11}
= - \frac{\chi v_{1}v_{2}v_{3}}{2} \int_{V_0}\text{d}q_{1} \text{d}q_{2} \text{d}q_{3} \frac{v^2_{2}q^{2}_{2}+v^{2}_{3}q^{2}_{3}-2v^{2}_{1}q^{2}_{1}}{\left(v^{2}_{1}q^{2}_{1}+v^{2}_{2}q^{2}_{2}+v^{2}_{3}q^{2}_{3}\right)^{\frac{5}{2}}},
\end{aligned}
\end{equation}
\\
outside the~volume $V_{0}$ enclosed by the aforementioned ellipsoid. $\chi$ is defined for the lower band. Writing $p_{i}=v_{i}q_{i}$, we obtain
\begin{equation}
\begin{aligned}
D_{11}
= \mp \frac{\chi}{2} \int_{V_0} \text{d}p_{1} \text{d}p_{2} \text{d}p_{3}\frac{p^{2}_{2}+p^{2}_{3}-2p^{2}_{1}}{\left(p^{2}_{1}+p^{2}_{2}+p^{2}_{3}\right)^{\frac{5}{2}}} .
\end{aligned}
\end{equation}
To deal with the degeneracy at $\mathbf{p}=\mathbf{0}$, we note that for a sphere centered at the Weyl node in these coordinates ($p_{1}, p_{2}, p_{3}$) the integral vanishes regardless of its radius. Therefore, we can remove the largest sphere centered at the origin within this ellipsoid from the volume $V_0$ integrated over.~A~similar \mbox{procedure} can be carried out for type-II Weyl nodes upon noting that a~sphere around the Weyl node can be found such that both bands are occupied and thus can be neglected.

Notably, we have ${\int \frac{\text{d}\mathbf{k}}{(2\pi)^3}f_{\mathbf{k}}(\partial_{1}\Omega_{1}+\partial_{2}\Omega_{2}+\partial_{3}\Omega_{3})=0}$. Thus, for~the models considered above, we expect two of the Berry curvature dipoles $D_{11}, D_{22}$ and $D_{33}$ to be equal, with~the remaining component being equal to twice the former but with an opposite sign.

In Fig.~\ref{fig:8}, we demonstrate the nonlinear Hall effects by plotting the diagonal terms of $D_{ij}$ for the 2-in-2-out configuration, for which ($q_{1}, q_{2}, q_{3}$) coincide with the Cartesian \mbox{coordinates}, for a range of values of $\phi$ and $\theta$. We find that the \mbox{nonlinear} Hall effects vary with the strength of the $\mathcal{P}$-symmetry-breaking perturbations, which is consistent with the changing contributions of the tilted Weyl nodes.

We conclude by noting that the observation of a~significant contribution to the nonlinear Hall conductivity from the off-diagonal $D_{ij}$ components could suggest that the corresponding $\mathcal{C}_{n}$-symmetry is broken in a FTOM crystal. 

\bibliography{refs}

@article{Ramirez1994,
  doi     = {10.1146/annurev.ms.24.080194.002321},
  url     = {https://doi.org/10.1146/annurev.ms.24.080194.002321},
  author  = {Ramirez, A. P.},
  title   = {Strongly Geometrically Frustrated Magnets},
  journal = {Annu. Rev. Mater. Sci.},
  volume  = {24},
  pages   = {453--480},
  year    = {1994}
}

@book{Lacroix2011,
  doi     = {10.1007/978-3-642-10589-0},
  url     = {https://doi.org/10.1007/978-3-642-10589-0},
  editor    = {Lacroix, Claudine and Mendels, Philippe and Mila, Fr{\'e}d{\'e}ric},
  title     = {Introduction to Frustrated Magnetism: Materials, Experiments, Theory},
  publisher = {Springer},
  series    = {Springer Series in Solid-State Sciences},
  volume    = {164},
  year      = {2011}
}

@article{Moessner2006,
  doi     = {10.1063/1.2186278},
  url     = {https://doi.org/10.1063/1.2186278},
  author  = {Moessner, Roderich and Ramirez, Arthur P.},
  title   = {Geometrical frustration},
  journal = {Physics Today},
  volume  = {59},
  number  = {2},
  pages   = {24--29},
  year    = {2006}
}

@article{Castelnovo2008,
  doi     = {10.1038/nature06433},
  url     = {https://doi.org/10.1038/nature06433},
  author  = {Castelnovo, C. and Moessner, R. and Sondhi, S. L.},
  title   = {Magnetic monopoles in spin ice},
  journal = {Nature},
  volume  = {451},
  pages   = {42--45},
  year    = {2008}
}

@article{Jaubert2009,
  doi     = {10.1038/nphys1227},
  url     = {https://doi.org/10.1038/nphys1227},
  author  = {Jaubert, L. D. C. and Holdsworth, P. C. W.},
  title   = {Signature of magnetic monopole and {Dirac} string dynamics in spin ice},
  journal = {Nat. Phys.},
  volume  = {5},
  pages   = {258--261},
  year    = {2009}
}

@article{Castelnovo2012,
  doi     = {10.1146/annurev-conmatphys-020911-125058},
  url     = {https://doi.org/10.1146/annurev-conmatphys-020911-125058},
  author  = {Castelnovo, Claudio and Moessner, Roderich and Sondhi, Shivaji L.},
  title   = {Spin Ice, Fractionalization, and Topological Order},
  journal = {Annu. Rev. Condens. Matter Phys.},
  volume  = {3},
  pages   = {35--55},
  year    = {2012}
}

@article{Fennell2009,
  doi     = {10.1126/science.1177582},
  url     = {https://doi.org/10.1126/science.1177582},
  author  = {Fennell, T. and Deen, P. P. and Wildes, A. R. and Schmalzl, K. and Prabhakaran, D. and Boothroyd, A. T. and Aldus, R. J. and McMorrow, D. F. and Bramwell, S. T.},
  title   = {Magnetic {Coulomb} Phase in the Spin Ice {Ho$_2$Ti$_2$O$_7$}},
  journal = {Science},
  volume  = {326},
  pages   = {415--417},
  year    = {2009}
}

@article{Morris2009,
  doi     = {10.1126/science.1178868},
  url     = {https://doi.org/10.1126/science.1178868},
  author  = {Morris, D. J. P. and Tennant, D. A. and Grigera, S. A. and Klemke, B. and Castelnovo, C. and Moessner, R. and Czternasty, C. and Meissner, M. and Rule, K. C. and Hoffmann, J.-U. and Kiefer, K. and Gerischer, S. and Slobinsky, D. and Perry, R. S.},
  title   = {Dirac Strings and Magnetic Monopoles in the Spin Ice {Dy$_2$Ti$_2$O$_7$}},
  journal = {Science},
  volume  = {326},
  pages   = {411--414},
  year    = {2009}
}

@article{Bramwell2009,
  doi     = {10.1038/nature08500},
  url     = {https://doi.org/10.1038/nature08500},
  author  = {Bramwell, S. T. and Giblin, S. R. and Calder, S. and Aldus, R. and Prabhakaran, D. and Fennell, T.},
  title   = {Measurement of the charge and current of magnetic monopoles in spin ice},
  journal = {Nature},
  volume  = {461},
  pages   = {956--959},
  year    = {2009}
}

@article{Kadowaki2009,
  doi     = {10.1143/JPSJ.78.103706},
  url     = {https://doi.org/10.1143/JPSJ.78.103706},
  author  = {Kadowaki, Hiroaki and Doi, Naohiro and Aoki, Yuji and Tabata, Yoshikazu and Sato, Taku J. and Lynn, Jeffrey W. and Matsuhira, Kazuyuki and Hiroi, Zenji},
  title   = {Observation of Magnetic Monopoles in Spin Ice},
  journal = {J. Phys. Soc. Jpn.},
  volume  = {78},
  pages   = {103706},
  year    = {2009}
}

@article{Balents2010,
  doi     = {10.1038/nature08917},
  url     = {https://doi.org/10.1038/nature08917},
  author  = {Balents, Leon},
  title   = {Spin liquids in frustrated magnets},
  journal = {Nature},
  volume  = {464},
  pages   = {199--208},
  year    = {2010}
}

@article{Savary2017,
  doi     = {10.1088/0034-4885/80/1/016502},
  url     = {https://doi.org/10.1088/0034-4885/80/1/016502},
  author  = {Savary, Lucile and Balents, Leon},
  title   = {Quantum spin liquids: a review},
  journal = {Rep. Prog. Phys.},
  volume  = {80},
  pages   = {016502},
  year    = {2017}
}

@article{Knolle2019,
  doi     = {10.1146/annurev-conmatphys-031218-013401},
  url     = {https://doi.org/10.1146/annurev-conmatphys-031218-013401},
  author  = {Knolle, Johannes and Moessner, Roderich},
  title   = {A Field Guide to Spin Liquids},
  journal = {Annu. Rev. Condens. Matter Phys.},
  volume  = {10},
  pages   = {451--472},
  year    = {2019}
}

@article{Kivelson1987,
  doi     = {10.1103/PhysRevB.35.8865},
  url     = {https://doi.org/10.1103/PhysRevB.35.8865},
  author  = {Kivelson, Steven A. and Rokhsar, Daniel S. and Sethna, James P.},
  title   = {Topology of the resonating valence-bond state: Solitons and high-{$T_c$} superconductivity},
  journal = {Phys. Rev. B},
  volume  = {35},
  pages   = {8865--8868},
  year    = {1987}
}

@article{Pauling1935,
  doi     = {10.1021/ja01315a102},
  url     = {https://doi.org/10.1021/ja01315a102},
  author  = {Pauling, Linus},
  title   = {The Structure and Entropy of Ice and of Other Crystals with Some Randomness of Atomic Arrangement},
  journal = {J. Am. Chem. Soc.},
  volume  = {57},
  pages   = {2680--2684},
  year    = {1935}
}

@article{Harris1997,
  doi     = {10.1103/PhysRevLett.79.2554},
  url     = {https://doi.org/10.1103/PhysRevLett.79.2554},
  author  = {Harris, M. J. and Bramwell, S. T. and McMorrow, D. F. and Zeiske, T. and Godfrey, K. W.},
  title   = {Geometrical Frustration in the Ferromagnetic Pyrochlore {Ho$_2$Ti$_2$O$_7$}},
  journal = {Phys. Rev. Lett.},
  volume  = {79},
  pages   = {2554--2557},
  year    = {1997}
}

@article{Ramirez1999,
  doi     = {10.1038/20619},
  url     = {https://doi.org/10.1038/20619},
  author  = {Ramirez, A. P. and Hayashi, A. and Cava, R. J. and Siddharthan, R. and Shastry, B. S.},
  title   = {Zero-point entropy in `spin ice'},
  journal = {Nature},
  volume  = {399},
  pages   = {333--335},
  year    = {1999}
}

@article{Bramwell2001,
  doi     = {10.1126/science.1064761},
  url     = {https://doi.org/10.1126/science.1064761},
  author  = {Bramwell, Steven T. and Gingras, Michel J. P.},
  title   = {Spin Ice State in Frustrated Magnetic Pyrochlore Materials},
  journal = {Science},
  volume  = {294},
  pages   = {1495--1501},
  year    = {2001}
}

@article{Wan2011,
  title   = {{Topological semimetal and Fermi-arc surface states in the electronic structure of pyrochlore iridates}},
  author  = {Wan, Xiangang and Turner, Ari M. and Vishwanath, Ashvin and Savrasov, Sergey Y.},
  journal = {Phys. Rev. B},
  volume  = {83},
  number  = {20},
  pages   = {205101},
  year    = {2011},
  doi     = {10.1103/PhysRevB.83.205101},
  url     = {https://doi.org/10.1103/PhysRevB.83.205101}
}

@article{Berke2018,
doi = {10.1088/1367-2630/aab881},
url = {https://doi.org/10.1088/1367-2630/aab881},
year = {2018},
month = {apr},
publisher = {IOP Publishing},
volume = {20},
number = {4},
pages = {043057},
author = {Berke, Christoph and Michetti, Paolo and Timm, Carsten},
title = {{Stability of the Weyl-semimetal phase on the pyrochlore lattice}},
journal = {New J. Phys.}
}

@article{Halperin1987,
  title   = {Possible states for a three-dimensional electron gas in a strong magnetic field},
  author  = {Halperin, Bertrand I.},
  journal = {Jpn. J. Appl. Phys.},
  volume  = {26},
  number  = {S3-3},
  pages   = {1913--1919},
  year    = {1987},
  doi     = {10.7567/JJAPS.26S3.1913}
}

@article{Burkov2011,
  title   = {Weyl semimetal in a topological insulator multilayer},
  author  = {Burkov, A. A. and Balents, Leon},
  journal = {Phys. Rev. Lett.},
  volume  = {107},
  issue   = {12},
  pages   = {127205},
  year    = {2011},
  doi     = {10.1103/PhysRevLett.107.127205}
}

@article{Khomskii2014,
  doi     = {10.1038/ncomms5793},
  url     = {https://doi.org/10.1038/ncomms5793},
  author  = {Khomskii, D. I.},
  title   = {Magnetic monopoles and unusual dynamics of magnetoelectrics},
  journal = {Nat. Commun.},
  volume  = {5},
  pages   = {4793},
  year    = {2014}
}

@article{Franz2018Antichiral,
  title = {{Antichiral Edge States in a Modified Haldane Nanoribbon}},
  author = {Colom\'es, E. and Franz, M.},
  journal = {Phys. Rev. Lett.},
  volume = {120},
  issue = {8},
  pages = {086603},
  numpages = {5},
  year = {2018},
  month = {Feb},
  publisher = {American Physical Society},
  doi = {10.1103/PhysRevLett.120.086603},
  url = {https://link.aps.org/doi/10.1103/PhysRevLett.120.086603}
}

@article{Dubovik1990,
  title   = {Toroid moments in electrodynamics and solid-state physics},
  author  = {Dubovik, V. M. and Tugushev, V. V.},
  journal = {Phys. Rep.},
  volume  = {187},
  number  = {4},
  pages   = {145--202},
  year    = {1990},
  doi     = {10.1016/0370-1573(90)90042-Z}
}

@article{Gao2018Spin,
  title   = {Microscopic theory of spin toroidization in periodic crystals},
  author  = {Gao, Yang and Vanderbilt, David and Xiao, Di},
  journal = {Phys. Rev. B},
  volume  = {97},
  issue   = {13},
  pages   = {134423},
  year    = {2018},
  month   = {Apr},
  doi     = {10.1103/PhysRevB.97.134423},
  url     = {https://doi.org/10.1103/PhysRevB.97.134423},
  eprint  = {},
  archivePrefix = {arXiv}
}

@article{Vazifeh2013,
  title = {{Electromagnetic Response of Weyl Semimetals}},
  author = {Vazifeh, M. M. and Franz, M.},
  journal = {Phys. Rev. Lett.},
  volume = {111},
  issue = {2},
  pages = {027201},
  numpages = {5},
  year = {2013},
  month = {Jul},
  publisher = {American Physical Society},
  doi = {10.1103/PhysRevLett.111.027201},
  url = {https://link.aps.org/doi/10.1103/PhysRevLett.111.027201}
}

@article{Shitade2018,
  title   = {Theory of orbital magnetic quadrupole moment and magnetoelectric susceptibility},
  author  = {Shitade, Atsuo and Watanabe, Hikaru and Yanase, Youichi},
  journal = {Phys. Rev. B},
  volume  = {98},
  issue   = {2},
  pages   = {020407},
  year    = {2018},
  month   = {Jul},
  doi     = {10.1103/PhysRevB.98.020407},
  url     = {https://link.aps.org/doi/10.1103/PhysRevB.98.020407},
  note    = {{Rapid Communication}}
}

@article{Wu2019,
  title   = {Non-{A}belian band topology in noninteracting metals},
  author  = {Wu, QuanSheng and Soluyanov, Alexey A. and Bzdu{\v{s}}ek, Tom{\'a}{\v{s}}},
  journal = {Science},
  volume  = {365},
  number  = {6459},
  pages   = {1273--1277},
  year    = {2019},
  doi     = {10.1126/science.aau8740},
  url     = {https://doi.org/10.1126/science.aau8740}
}

@article{Ceresoli2006,
  title   = {Orbital magnetization in crystalline solids: Multi-band insulators, Chern insulators, and metals},
  author  = {Ceresoli, Davide and Thonhauser, T. and Vanderbilt, David and Resta, R.},
  journal = {Phys. Rev. B},
  volume  = {74},
  issue   = {2},
  pages   = {024408},
  year    = {2006},
  doi     = {10.1103/PhysRevB.74.024408}
}

@article{Xiao2010,
  title   = {Berry phase effects on electronic properties},
  author  = {Xiao, Di and Chang, Ming-Che and Niu, Qian},
  journal = {Rev. Mod. Phys.},
  volume  = {82},
  issue   = {3},
  pages   = {1959--2007},
  year    = {2010},
  month   = {Jul},
  doi     = {10.1103/RevModPhys.82.1959},
  url     = {https://doi.org/10.1103/RevModPhys.82.1959}
}

@article{Tiwari2020,
  title   = {Non-{A}belian topology of nodal-line rings in $\mathcal{PT}$-symmetric systems},
  author  = {Tiwari, Apoorv and Bzdu{\v{s}}ek, Tom{\'a}{\v{s}}},
  journal = {Phys. Rev. B},
  volume  = {101},
  issue   = {19},
  pages   = {195130},
  year    = {2020},
  month   = {May},
  doi     = {10.1103/PhysRevB.101.195130},
  url     = {https://doi.org/10.1103/PhysRevB.101.195130},
  eprint  = {},
  archivePrefix = {arXiv}
}

@article{Gao2018,
  title   = {Orbital magnetic quadrupole moment and nonlinear anomalous thermoelectric transport},
  author  = {Gao, Yang and Xiao, Di},
  journal = {Phys. Rev. B},
  volume  = {98},
  issue   = {6},
  pages   = {060402},
  year    = {2018},
  month   = {Aug},
  doi     = {10.1103/PhysRevB.98.060402},
  url     = {https://link.aps.org/doi/10.1103/PhysRevB.98.060402},
  note    = {{Rapid Communication}}
}

@article{Haldane1988,
  title = {{Model for a Quantum Hall Effect without Landau Levels: Condensed-Matter Realization of the "Parity Anomaly"}},
  author = {Haldane, F. D. M.},
  journal = {Phys. Rev. Lett.},
  volume = {61},
  issue = {18},
  pages = {2015--2018},
  numpages = {0},
  year = {1988},
  month = {Oct},
  publisher = {American Physical Society},
  doi = {10.1103/PhysRevLett.61.2015},
  url = {https://link.aps.org/doi/10.1103/PhysRevLett.61.2015}
}

@article{BzdusekTP,
  title = {Triple nodal points characterized by their nodal-line structure in all magnetic space groups},
  author = {Lenggenhager, Patrick M. and Liu, Xiaoxiong and Neupert, Titus and Bzdu\ifmmode \check{s}\else \v{s}\fi{}ek, Tom\'a\ifmmode \check{s}\else \v{s}\fi{}},
  journal = {Phys. Rev. B},
  volume = {106},
  issue = {8},
  pages = {085128},
  numpages = {36},
  year = {2022},
  month = {Aug},
  publisher = {American Physical Society},
  doi = {10.1103/PhysRevB.106.085128},
  url = {https://link.aps.org/doi/10.1103/PhysRevB.106.085128}
}

@article{GaoQ,
  title = {Orbital magnetic quadrupole moment and nonlinear anomalous thermoelectric transport},
  author = {Gao, Yang and Xiao, Di},
  journal = {Phys. Rev. B},
  volume = {98},
  issue = {6},
  pages = {060402(R)},
  numpages = {5},
  year = {2018},
  month = {Aug},
  publisher = {American Physical Society},
  doi = {10.1103/PhysRevB.98.060402},
  url = {https://link.aps.org/doi/10.1103/PhysRevB.98.060402}
}

@misc{Mineev2025,
      title={Toroid, Altermagnetic and Noncentrosymmetric ordering in metals}, 
      author={V. P. Mineev},
      year={2026},
      eprint={2504.01686},
      archivePrefix={arXiv},
      primaryClass={cond-mat.str-el},
      url={https://arxiv.org/abs/2504.01686}, 
}

@article{Resta2006,
  title = {Orbital magnetization in crystalline solids: Multi-band insulators, Chern insulators, and metals},
  author = {Ceresoli, Davide and Thonhauser, T. and Vanderbilt, David and Resta, R.},
  journal = {Phys. Rev. B},
  volume = {74},
  issue = {2},
  pages = {024408},
  numpages = {13},
  year = {2006},
  month = {Jul},
  publisher = {American Physical Society},
  doi = {10.1103/PhysRevB.74.024408},
  url = {https://link.aps.org/doi/10.1103/PhysRevB.74.024408}
}

@article{Armitage2018,
  title   = {Weyl and Dirac semimetals in three-dimensional solids},
  author  = {Armitage, N. P. and Mele, E. J. and Vishwanath, Ashvin},
  journal = {Rev. Mod. Phys.},
  volume  = {90},
  issue   = {1},
  pages   = {015001},
  year    = {2018},
  month   = {Jan},
  doi     = {10.1103/RevModPhys.90.015001},
  url     = {https://doi.org/10.1103/RevModPhys.90.015001}
}

@misc{Yamanaka2026,
      title={Magnetic toroidal monopoles from relativistic polarization responses to magnetic field gradients}, 
      author={Taisei Yamanaka and Takumi Sato and Satoru Hayami},
      year={2026},
      eprint={2604.05412},
      archivePrefix={arXiv},
      primaryClass={cond-mat.str-el},
      url={https://arxiv.org/abs/2604.05412}, 
}

@unpublished{ZexunPriv,
  author = {Lin, Ze-Xun},
  title  = {},
  date   = {2026-07-16},
  note   = {Private communication}
}

@article{deJuan2017,
  title   = {Quantized circular photogalvanic effect in {W}eyl semimetals},
  author  = {de Juan, Fernando and Grushin, Adolfo G. and Morimoto, Takahiro and Moore, Joel E.},
  journal = {Nat. Commun.},
  volume  = {8},
  pages   = {15995},
  year    = {2017},
  doi     = {10.1038/ncomms15995},
  url     = {https://doi.org/10.1038/ncomms15995}
}

@article{Spaldin2008,
  author  = {Spaldin, Nicola A. and Fiebig, Manfred and Mostovoy, Maxim},
  title   = {The toroidal moment in condensed-matter physics and its relation to the magnetoelectric effect},
  journal = {J. Phys. Condens. Matter},
  volume  = {20},
  number  = {43},
  pages   = {434203},
  year    = {2008},
  doi     = {10.1088/0953-8984/20/43/434203},
  publisher = {IOP Publishing}
}

@article{Witczak2013,
  title = {Pyrochlore electrons under pressure, heat, and field: Shedding light on the iridates},
  author = {Witczak-Krempa, William and Go, Ara and Kim, Yong Baek},
  journal = {Phys. Rev. B},
  volume = {87},
  issue = {15},
  pages = {155101},
  numpages = {7},
  year = {2013},
  month = {Apr},
  publisher = {American Physical Society},
  doi = {10.1103/PhysRevB.87.155101},
  url = {https://link.aps.org/doi/10.1103/PhysRevB.87.155101}
}

@article{Shinaoka2015,
  title   = {{Phase Diagram of Pyrochlore Iridates: All-in--All-out Magnetic Ordering and Non-Fermi-Liquid Properties}},
  author  = {Shinaoka, Hiroshi and Hoshino, Shintaro and Troyer, Matthias and Werner, Philipp},
  journal = {Phys. Rev. Lett.},
  volume  = {115},
  issue   = {15},
  pages   = {156401},
  year    = {2015},
  month   = {Oct},
  publisher = {American Physical Society},
  doi     = {10.1103/PhysRevLett.115.156401},
  url     = {https://link.aps.org/doi/10.1103/PhysRevLett.115.156401}
}

@article{Zhang2017,
  title     = {{Metal-Insulator Transition and Topological Properties of Pyrochlore Iridates}},
  author    = {Zhang, Hongbin and Haule, Kristjan and Vanderbilt, David},
  journal   = {Phys. Rev. Lett.},
  volume    = {118},
  issue     = {2},
  pages     = {026404},
  year      = {2017},
  month     = {Jan},
  publisher = {American Physical Society},
  doi       = {10.1103/PhysRevLett.118.026404},
  url       = {https://link.aps.org/doi/10.1103/PhysRevLett.118.026404}
}

@article{ChenLee2011,
  title   = {Unified formalism for calculating polarization, magnetization, and more in a periodic insulator},
  author  = {Chen, Kuang-Ting and Lee, Patrick A.},
  journal = {Phys. Rev. B},
  volume  = {84},
  issue   = {20},
  pages   = {205137},
  year    = {2011},
  month   = nov,
  publisher = {American Physical Society},
  doi     = {10.1103/PhysRevB.84.205137},
  url     = {https://doi.org/10.1103/PhysRevB.84.205137},
  eprint  = {},
  archivePrefix = {arXiv},
  primaryClass  = {cond-mat.mes-hall}
}

@article{Soluyanov2015,
  title     = {{Type-II Weyl semimetals}},
  author    = {Soluyanov, Alexey A. and Gresch, Dominik and Wang, Zhijun and Wu, QuanSheng and Troyer, Matthias and Dai, Xi and Bernevig, B. Andrei},
  journal   = {Nature},
  volume    = {527},
  number    = {7579},
  pages     = {495--498},
  year      = {2015},
  publisher = {Nature Publishing Group},
  doi       = {10.1038/nature15768}
}

@article{Gilbert2021,
  title   = {Topological electronics},
  author  = {Gilbert, Matthew J.},
  journal = {Commun. Phys.},
  volume  = {4},
  pages   = {70},
  year    = {2021},
  doi     = {10.1038/s42005-021-00569-5}
}

@article{He2022,
  title   = {Topological spintronics and magnetoelectronics},
  author  = {He, Qing Lin and Hughes, Taylor L. and Armitage, N. Peter and Tokura, Yoshinori and Wang, Kang L.},
  journal = {Nat. Mater.},
  volume  = {21},
  pages   = {15--23},
  year    = {2022},
  doi     = {10.1038/s41563-021-01138-5}
}

@article{Levin2005,
  author  = {Levin, Michael and Wen, Xiao-Gang},
  title   = {Colloquium: Photons and electrons as emergent phenomena},
  journal = {Rev. Mod. Phys.},
  volume  = {77},
  pages   = {871--879},
  year    = {2005},
  doi     = {10.1103/RevModPhys.77.871}
}

@book{Wen2004,
  author    = {Wen, Xiao-Gang},
  title     = {Quantum Field Theory of Many-Body Systems: From the Origin of Sound to an Origin of Light and Electrons},
  publisher = {Oxford University Press},
  address   = {Oxford},
  year      = {2004},
  url = {https://academic.oup.com/book/25836}
}

@article{Broholm2020,
  author  = {Broholm, C. and Cava, R. J. and Kivelson, S. A. and Nocera, D. G. and Norman, M. R. and Senthil, T.},
  title   = {Quantum spin liquids},
  journal = {Science},
  volume  = {367},
  number  = {6475},
  pages   = {eaay0668},
  year    = {2020},
  doi     = {10.1126/science.aay0668}
}

@article{Wen2017,
  author  = {Wen, Xiao-Gang},
  title   = {Colloquium: Zoo of quantum-topological phases of matter},
  journal = {Rev. Mod. Phys.},
  volume  = {89},
  pages   = {041004},
  year    = {2017},
  doi     = {10.1103/RevModPhys.89.041004}
}

@book{Udagawa2021,
  editor    = {Udagawa, Masafumi and Jaubert, Ludovic},
  title     = {Spin Ice},
  series    = {Springer Series in Solid-State Sciences},
  volume    = {197},
  publisher = {Springer},
  address   = {Cham},
  year      = {2021},
  isbn      = {978-3-030-70860-3},
  doi       = {10.1007/978-3-030-70860-3},
  url       = {https://doi.org/10.1007/978-3-030-70860-3}
}

@article{Sodemann2015,
  title = {{Quantum Nonlinear Hall Effect Induced by Berry Curvature Dipole in Time-Reversal Invariant Materials}},
  author = {Sodemann, Inti and Fu, Liang},
  journal = {Phys. Rev. Lett.},
  volume = {115},
  issue = {21},
  pages = {216806},
  numpages = {5},
  year = {2015},
  month = {Nov},
  publisher = {American Physical Society},
  doi = {10.1103/PhysRevLett.115.216806},
  url = {https://link.aps.org/doi/10.1103/PhysRevLett.115.216806}
}

\end{document}